\documentclass[aps,prd,twocolumn,superscriptaddress,nofootinbib,longbibliography,preprintnumbers]{revtex4-1}

\usepackage{amsmath,amssymb,amsthm,bm,graphicx,xcolor,mathrsfs}
\usepackage[caption=false]{subfig}
\usepackage[colorlinks=true,linkcolor=blue,citecolor=blue,urlcolor=blue]{hyperref}

\begin{document}

\preprint{NU-QG-25}

\title{Long-term 3+1 simulations of primordial black hole \\formation during radiation domination}


\author{Zhuan Ning}
\email{ningzhuan17@mails.ucas.ac.cn}
\affiliation{School of Fundamental Physics and Mathematical Sciences, Hangzhou Institute for Advanced Study, UCAS, Hangzhou 310024, China}
\affiliation{Institute of Theoretical Physics, Chinese Academy of Sciences, Beijing 100190, China}
\affiliation{University of Chinese Academy of Sciences, Beijing 100049, China}

\author{Rong-Gen Cai}
\email{caironggen@nbu.edu.cn}
\affiliation{Institute of Fundamental Physics and Quantum Technology, \& School of Physical Science and Technology, Ningbo University, Ningbo 315211, China}

\author{Shao-Jiang Wang}
\email{schwang@itp.ac.cn}
\affiliation{Institute of Theoretical Physics, Chinese Academy of Sciences, Beijing 100190, China}
\affiliation{Asia Pacific Center for Theoretical Physics (APCTP), Pohang 37673, Korea}

\author{Chul-Moon Yoo}
\email{yoo.chulmoon.k6@f.mail.nagoya-u.ac.jp}
\affiliation{Graduate School of Science, Nagoya University, Nagoya 464-8602, Japan}
\affiliation{Kobayashi-Maskawa Institute for the Origin of Particles and the Universe (KMI), Nagoya 464-8602, Japan}

\begin{abstract}
We develop an efficient three-dimensional numerical-relativity framework for primordial black-hole (PBH) formation from superhorizon curvature perturbations in a radiation-dominated Universe. We implement flux-conservative relativistic hydrodynamics in the adaptive-mesh-refinement code \textsc{GRChombo} and introduce a cosmologically scaled Gamma-driver that allows the cosmic-time step to grow in proportion to the scale factor. For a representative long-term simulation, the scaled driver preserves the apparent-horizon mass evolution and constraint behavior while reducing the number of coarse-level advances by a factor of approximately $94$ relative to the standard driver. We also construct a conformal-time version of the moving-puncture gauge as a consistency check. Applying the framework to a spherical Gaussian curvature profile, we find a collapse threshold $\mu_c \simeq 0.7958$ and approximate critical mass scaling, consistent with previous spherically symmetric results. We further fit the late-time PBH mass growth to the Zel'dovich--Novikov accretion law, demonstrating that the code can follow both near-critical collapse and long-term post-formation evolution in three dimensions. The framework provides a foundation for future studies of PBH formation beyond spherical symmetry.
\end{abstract}

\maketitle

\section{Introduction}

Primordial black holes (PBHs) are black holes that may have formed in the early Universe rather than through the late-time collapse of stars~\cite{Zeldovich:1967lct,Hawking:1971ei,Carr:1974nx,Carr:1975qj,Chapline:1975ojl}. Their existence would provide direct evidence for sufficiently large primordial inhomogeneities and offer a probe of the statistics of small-scale primordial perturbations~\cite{Bugaev:2010bb,Dong:2015yjs,Kimura:2021sqz,Wang:2022nml,Inui:2024fgk,Ning:2025ogq,Zeng:2025law,Zeng:2025cer,Farooq:2025qmm,Ning:2025yvj,Ding:2026tlj}. Depending on their abundance and mass distribution, PBHs may also contribute to dark matter~\cite{Carr:2009jm,Belotsky:2014kca,Carr:2016drx,Carr:2020gox,Carr:2020xqk,Green:2020jor,Carr:2026hot}, source gravitational-wave signals through binary mergers~\cite{Bird:2016dcv,Sasaki:2016jop,Sasaki:2018dmp}, and provide seeds for supermassive black holes~\cite{Bean:2002kx,Duechting:2004dk,Kawasaki:2012kn,Carr:2018rid}. These possibilities make their formation conditions and initial mass function central ingredients in connecting early-Universe physics with present-day observations.

Among the proposed formation mechanisms, a widely studied scenario is the gravitational collapse of large primordial curvature perturbations during radiation domination, which is the primary focus of this work. Such perturbations remain approximately frozen while they are well outside the Hubble horizon and begin to evolve dynamically after horizon reentry. A PBH forms only when the perturbation is sufficiently large for self-gravity to overcome the pressure gradients of the radiation fluid. The corresponding collapse threshold depends on the equation of state~\cite{Harada:2013epa,Escriva:2020tak} and on the shape of the initial curvature profile~\cite{Musco:2018rwt,Escriva:2019phb}, while perturbations immediately above threshold exhibit critical mass scaling~\cite{Niemeyer:1999ak,Hawke:2002rf,Musco:2004ak,Musco:2008hv,Musco:2012au}.

Accurate determination of the collapse threshold is particularly important because the predicted PBH abundance is exponentially sensitive to the formation criterion~\cite{Carr:1975qj}. Relativistic numerical simulations are generally required to follow the nonlinear evolution through horizon reentry and black-hole formation. Most calculations have imposed spherical symmetry, for which the collapse threshold, critical behavior, and subsequent mass growth have been studied extensively~\cite{Shibata:1999zs,Niemeyer:1999ak,Hawke:2002rf,Musco:2004ak,Musco:2008hv,Musco:2012au,Deng:2016vzb,Musco:2018rwt,Escriva:2019nsa,Joana:2025gqf,Ning:2026nfs,Yuwen:2026hxu,Escriva:2026hel} (see Ref.~\cite{Escriva:2021aeh} for a review). In contrast, fully three-dimensional simulations remain much less common, with only a few studies achieving this goal~\cite{Yoo:2020lmg,Yoo:2024lhp,Escriva:2024lmm,Escriva:2024aeo,Baumgarte:2026igz,Escriva:2026kov}.\footnote{See also Refs.~\cite{deJong:2021bbo,deJong:2023gsx,Cheng:2025eas,Kitajima:2025shn,Adshead:2025gka,Yoo:2026frk,Munoz:2026pce} for three-dimensional numerical-relativity simulations of PBH formation through other mechanisms.} Establishing an efficient three-dimensional framework is nevertheless essential for studying the effects of nonsphericity in realistic primordial perturbations and the spin acquired by PBHs.

PBH formation poses a demanding multiscale problem. A simulation must encompass an initially superhorizon curvature perturbation while resolving the much smaller local length scales associated with collapse after horizon entry. This separation becomes more severe near criticality, where the newly formed black-hole mass decreases according to the critical scaling law. The \textsc{COSMOS} code mitigates this spatial cost through scale-up coordinates and fixed mesh refinement~\cite{Yoo:2018pda,Yoo:2020lmg,Yoo:2024lhp,Escriva:2024lmm,Yoo:2026itl}. Adaptive mesh refinement (AMR) provides another natural way to concentrate resolution around the collapsing region. These techniques alleviate the spatial bottleneck, but they do not reduce the number of time steps required to follow post-formation PBH evolution over many Hubble times. The remaining challenge is therefore temporal.

Whether this temporal cost can be reduced depends crucially on the gauge conditions. Several existing three-dimensional studies employ cosmological modifications of the moving-puncture gauge, combining a background-subtracted $1 + \log$ slicing condition with a Gamma-driver shift~\cite{Yoo:2020lmg,Yoo:2024lhp,Escriva:2024lmm}. This construction retains the singularity-avoidance mechanism that makes the moving-puncture gauge effective for asymptotically flat black-hole spacetimes~\cite{Alcubierre:2002kk,vanMeter:2006vi,Ning:2026qxs}, while the coordinate time approaches cosmic time in the weakly perturbed FLRW region. As the Universe expands, the coordinate speeds of physical modes decline inversely with the scale factor, as $a^{-1}$, making a fixed time step increasingly conservative relative to the physical Courant--Friedrichs--Lewy (CFL) limit. In principle, this limit would permit the time step $\Delta t$ to grow in proportion to the scale factor. However, the characteristic speeds and damping timescale of the standard Gamma-driver prevent such a scaling, as we demonstrate in Sec.~\ref{sec:cosmological-gauges}.

In this work, we implement a flux-conservative relativistic-fluid evolution module in the AMR numerical-relativity code \textsc{GRChombo}~\cite{Clough:2015sqa,Radia:2021smk,Andrade:2021rbd} and construct a complete framework for simulating PBH formation from superhorizon curvature perturbations during radiation domination. AMR addresses the spatial scale separation by dynamically resolving the collapsing region. To accelerate the long-time evolution, we introduce a cosmologically scaled Gamma-driver, which permits $\Delta t \propto a\Delta x$, corresponding to an approximately constant conformal-time step while retaining cosmic time as the evolution coordinate. We also construct a conformal-time version of the moving-puncture gauge as a consistency and robustness check.

As a first application, we consider a spherical Gaussian curvature perturbation in a radiation-dominated Universe. The three-dimensional simulations yield a formation threshold consistent with both previous three-dimensional results and a spherically symmetric Misner--Sharp calculation for the same profile. The near-critical simulations exhibit approximate critical mass scaling, with the accuracy of the fitted exponent limited by the accessible black-hole mass range and numerical resolution. We further evolve the newly formed PBHs over many Hubble times and fit their late-time mass growth to the Zel'dovich--Novikov accretion prescription~\cite{Zeldovich:1967lct}. To the best of our knowledge, these are the first three-dimensional PBH formation simulations used to extract both the critical mass scaling relation and the post-formation accretion law.

The remainder of this paper is organized as follows. In Sec.~\ref{sec:formulation}, we summarize the BSSN and relativistic-hydrodynamics formulations, construct the long-wavelength initial data, and describe the numerical setup. In Sec.~\ref{sec:cosmological-gauges}, we analyze the cosmological gauge conditions and introduce the scaled Gamma-driver and conformal-time gauge. Section~\ref{sec:results} presents the gauge comparisons, boundary-condition tests, collapse threshold, critical mass scaling, and post-formation mass growth. We conclude and discuss future applications in Sec.~\ref{sec:conclusions}. Appendices~\ref{app:fluid} and~\ref{app:convergence} summarize the fluid evolution scheme and the numerical tests, respectively. Throughout this work, we use geometrized units with $c = G = 1$. Greek indices run over spacetime coordinates $(0, 1, 2, 3)$, whereas Latin indices run over spatial coordinates $(1, 2, 3)$.

\section{Formulation and numerical setup}\label{sec:formulation}

\subsection{Spacetime evolution}

We write the spacetime metric in the standard $3+1$ form~\cite{Gourgoulhon:2007ue,2008itnr.book.....A,2016nure.book.....S},
\begin{equation}
    \mathrm{d}s^2 = -\alpha^2 \mathrm{d}t^2 + \gamma_{ij} \left(\mathrm{d}x^i + \beta^i \mathrm{d}t\right) \left(\mathrm{d}x^j + \beta^j \mathrm{d}t\right),
\end{equation}
where $\alpha$, $\beta^i$, and $\gamma_{ij}$ denote the lapse function, shift vector, and spatial metric, respectively. The future-directed unit vector normal to a constant-$t$ hypersurface is $n_\mu = (-\alpha, 0, 0, 0)$, and we define the extrinsic curvature as
\begin{equation}
    K_{ij} = -\frac{1}{2\alpha} \left(\partial_t \gamma_{ij} - \mathcal{L}_\beta \gamma_{ij}\right).
\end{equation}

We evolve the Einstein equations using the Baumgarte--Shapiro--Shibata--Nakamura (BSSN) formulation~\cite{Nakamura:1987zz,Shibata:1995we,Baumgarte:1998te}. The evolved geometric variables are
\begin{equation}
    \begin{aligned}
        \chi &\equiv \gamma^{-1/3}, & \tilde{\gamma}_{ij} &\equiv \chi \gamma_{ij}, \\
        K &\equiv \gamma^{ij} K_{ij}, & \tilde{A}_{ij} &\equiv \chi \left(K_{ij} - \frac{1}{3}\gamma_{ij} K\right), \\
        \tilde{\Gamma}^{i} &\equiv \tilde{\gamma}^{jk} \tilde{\Gamma}^{i}_{jk},
    \end{aligned}
\end{equation}
where $\gamma \equiv \det(\gamma_{ij})$. We do not reproduce the standard BSSN evolution equations here. The matter source terms entering these equations are given by the Eulerian projections
\begin{equation}
    \begin{aligned}
        E &\equiv n_\mu n_\nu T^{\mu\nu}, \\
        S_i &\equiv -\gamma_{i\mu} n_\nu T^{\mu\nu}, \\
        S_{ij} &\equiv \gamma_{i\mu} \gamma_{j\nu} T^{\mu\nu}, \\
        S &\equiv \gamma^{ij} S_{ij}.
    \end{aligned}
\end{equation}
The Hamiltonian and momentum constraints then take the form
\begin{subequations}
    \begin{align}
        \mathscr{H} &\equiv {}^{(3)}R + K^2 - K_{ij} K^{ij} - 16\pi E = 0, \\
        \mathscr{M}_i &\equiv D_j K^j{}_i - D_i K - 8\pi S_i = 0,
    \end{align}
    \label{eq:constraints}
\end{subequations}
where ${}^{(3)}R$ and $D_i$ denote the Ricci scalar and covariant derivative associated with $\gamma_{ij}$, respectively.

\subsection{Relativistic hydrodynamics}

For a radiation-dominated Universe, we describe the matter sector by a perfect fluid with the stress-energy tensor
\begin{equation}
    T^{\mu\nu} = (\rho + p)u^\mu u^\nu + p g^{\mu\nu},
\end{equation}
where $\rho$ is the proper energy density measured in the fluid rest frame, $p$ is the pressure, and $u^\mu$ is the fluid four-velocity. We adopt a linear barotropic equation of state,
\begin{equation}
    p = \omega \rho,
\end{equation}
and set $\omega = 1 / 3$ throughout this work to describe the radiation fluid. For this equation of state, we do not need to decompose $\rho$ into rest-mass and internal-energy contributions. Defining the Lorentz factor as $W \equiv -n_\mu u^\mu$, we decompose the fluid four-velocity as
\begin{equation}
    \begin{aligned}
        u^\mu &= W(n^\mu + v^\mu), \quad \text{with} \quad n_\mu v^\mu = 0, \\
        W &= (1 - v^2)^{-1/2},
    \end{aligned}
\end{equation}
where $v^2 \equiv \gamma_{ij} v^i v^j$. The Eulerian projections of the stress-energy tensor then become
\begin{equation}
    \begin{aligned}
        E &= (\rho + p)W^2 - p, \\
        S_i &= (\rho + p)W^2 v_i, \\
        S_{ij} &= (\rho + p)W^2 v_i v_j + p \gamma_{ij}.
    \end{aligned}
    \label{eq:fluid_projections}
\end{equation}

To evolve the fluid, we use a conservative formulation adapted from the \textsc{COSMOS} code~\cite{Yoo:2020lmg,Yoo:2024lhp,Escriva:2024lmm,Escriva:2024aeo,Yoo:2026itl}. Specifically, we evolve the densitized conserved variables
\begin{equation}
    \mathcal{E} \equiv \sqrt{\gamma}\,E, \quad \mathcal{S}_i \equiv \sqrt{\gamma}\,S_i.
\end{equation}
Energy-momentum conservation, $\nabla_\mu T^{\mu\nu} = 0$, then takes the flux-conservative form
\begin{subequations}
    \begin{align}
        \partial_t \mathcal{E} + \partial_j \left[\mathcal{E}(\alpha v^j - \beta^j) + \alpha\sqrt{\gamma}\,p v^j\right] &= \mathcal{Q}_{E}, \\
        \partial_t \mathcal{S}_i + \partial_j \left[\mathcal{S}_i(\alpha v^j - \beta^j) + \alpha\sqrt{\gamma}\,p\delta_i{}^j\right] &= \mathcal{Q}_{S_i},
    \end{align}
    \label{eq:fluid_conservative_form}
\end{subequations}
where
\begin{subequations}
    \begin{align}
        \mathcal{Q}_{E} &= \alpha\sqrt{\gamma}\,S^{ij}K_{ij} - \sqrt{\gamma}\,S^i \partial_i \alpha, \\
        \mathcal{Q}_{S_i} &= -\mathcal{E}\,\partial_i \alpha + \mathcal{S}_j \partial_i \beta^j + \frac{1}{2}\alpha\sqrt{\gamma}\,S^{jk} \partial_i \gamma_{jk}.
    \end{align}
\end{subequations}
The primitive variables $(\rho, v^i)$ are not evolved directly, but are recovered from the conserved variables $(\mathcal{E}, \mathcal{S}_i)$ after each time-integration substep. For a linear barotropic equation of state, the conserved-to-primitive conversion can be performed analytically. The fluid equations~\eqref{eq:fluid_conservative_form} are discretized using a monotonic upstream-centered scheme for conservation laws (MUSCL) reconstruction and a local Lax--Friedrichs flux~\cite{Kurganov:2000ovy,Shibata:2005jv}. Further details of the numerical scheme are provided in Appendix~\ref{app:fluid}.

\subsection{Initial data}

For the initial data, we construct a curvature perturbation well outside the Hubble horizon using the long-wavelength, or gradient-expansion, solution of Refs.~\cite{Shibata:1999zs,Harada:2015yda}. In terms of the curvature perturbation $\zeta(\boldsymbol{x})$, the spatial metric takes the form
\begin{equation}
    \gamma_{ij} = a_i^2 e^{2\zeta(\boldsymbol{x})}\delta_{ij} + \mathcal{O}(\epsilon^2), \quad \epsilon \equiv \frac{k}{a_i H_i} \ll 1,
\end{equation}
where $1/k$ gives the characteristic comoving scale of the perturbation, and $a_i$ and $H_i$ are the scale factor and Hubble rate on the initial slice, respectively. Without loss of generality, we set $a_i = 1$.

In this work, we consider a spherical Gaussian profile for the curvature perturbation,
\begin{equation}
    \zeta(r) = \mu \exp\left(-k^2 r^2\right) W(r),
    \label{eq:spherical_zeta}
\end{equation}
where $\mu$ is the amplitude of the perturbation, and $W(r)$ is a window function that smoothly brings the perturbation to zero at the outer boundaries without affecting the profile around its characteristic scale. We adopt the following window function~\cite{Yoo:2018pda,Yoo:2020lmg}:
\begin{equation}
    W(r) = \begin{cases}
        1, & 0 \leq r \leq r_W, \\
        \displaystyle 1 - \frac{\left[(L - r_W)^6 - (L - r)^6\right]^6}{(L - r_W)^{36}}, & r_W < r < L, \\
        0, & r \geq L.
    \end{cases}
\end{equation}
Here, $L$ is the half-side length of the cubic simulation domain, and $r_W$ denotes the radius at which the window begins to taper off. For the Gaussian profile in Eq.~\eqref{eq:spherical_zeta}, the leading-order compaction function reaches its maximum at $r_m = 1/k$, and we take $r_m$ as the characteristic scale of the perturbation~\cite{Musco:2018rwt}.\footnote{Our definition of $k$ is related to that used by Yoo et al. in Ref.~\cite{Yoo:2020lmg} through $k = k_\mathrm{Yoo}/\sqrt{6}$. Yoo et al. use $1/k_\mathrm{Yoo}$ as the characteristic scale, whereas our compaction-maximum definition gives $r_m = \sqrt{6}/k_\mathrm{Yoo} = 1/k$; the corresponding horizon-entry time and horizon-mass normalizations therefore differ.}

After specifying $\zeta(\boldsymbol{x})$, we evaluate the long-wavelength solution in constant-mean-curvature slicing with homogeneous $K$ and normal threading with a vanishing shift vector~\cite{Harada:2015yda}. Rewriting the result of Ref.~\cite{Harada:2015yda} in terms of $\zeta$, we first define the scalar and trace-free tensor
\begin{subequations}
    \begin{align}
        f &\equiv -\frac{2}{3} e^{-2\zeta}\left[\partial^2\zeta + \frac{1}{2}(\partial\zeta)^2\right], \\
        p_{ij} &\equiv -e^{-2\zeta}\left(\partial_i\partial_j\zeta - \frac{1}{3}\delta_{ij}\partial^2\zeta\right) \notag \\
        &\quad + e^{-2\zeta}\left(\partial_i\zeta\partial_j\zeta - \frac{1}{3}\delta_{ij}(\partial\zeta)^2\right),
    \end{align}
\end{subequations}
where $\partial^2 \equiv \delta^{ij}\partial_i\partial_j$ and $(\partial\zeta)^2 \equiv \delta^{ij}\partial_i\zeta\partial_j\zeta$. In our BSSN variables, the growing-mode solution through $\mathcal{O}(\epsilon^2)$ is
\begin{subequations}
    \begin{align}
        \chi &= a_i^{-2}e^{-2\zeta}\left[1 + \frac{2f}{3(1 + \omega)}\frac{1}{(a_i H_i)^2}\right] + \mathcal{O}(\epsilon^4), \\
        \tilde{\gamma}_{ij} &= \delta_{ij} - \frac{4p_{ij}}{(1 + 3\omega)(5 + 3\omega)}\frac{1}{(a_i H_i)^2} + \mathcal{O}(\epsilon^4), \\
        K &= -3H_i, \\
        \tilde{A}_{ij} &= \frac{2H_i p_{ij}}{5 + 3\omega}\frac{1}{(a_i H_i)^2} + \mathcal{O}(\epsilon^4), \\
        \alpha &= 1 - \frac{1 + 3\omega}{3(1 + \omega)}f\frac{1}{(a_i H_i)^2} + \mathcal{O}(\epsilon^4), \qquad \beta^i = 0.
    \end{align}
\end{subequations}
We use these expressions to construct the initial geometric and gauge variables. The fluid variables are then determined algebraically from the Hamiltonian and momentum constraint equations~\eqref{eq:constraints}, rather than from their truncated long-wavelength expressions~\cite{Yoo:2024lhp}.

For constant $\omega$, the reference FLRW solution can be written in terms of the cosmic time $t$ as
\begin{equation}
    a(t) = \left(\frac{t}{t_i}\right)^{\alpha_\omega}, \quad H(t) = \frac{\alpha_\omega}{t}, \quad \alpha_\omega \equiv \frac{2}{3(1 + \omega)}.
\end{equation}
Here, $t_i = \alpha_\omega / H_i$. We define horizon entry by $a(t_H) r_m = H^{-1}(t_H)$, which gives
\begin{equation}
    \begin{aligned}
        t_H &= t_i \left(a_i H_i r_m\right)^{1/(1 - \alpha_\omega)}, \\
        M_H &\equiv \frac{1}{2H(t_H)} = \frac{t_H}{2\alpha_\omega}.
    \end{aligned}
\end{equation}
We use $r_m$, $t_H$, and $M_H$ to normalize the radius, time, and black hole mass in Sec.~\ref{sec:results}, respectively.

Alternatively, we can express the reference FLRW solution in terms of conformal time $\eta$ as
\begin{equation}
    a(\eta) = \left(\frac{\eta}{\eta_i}\right)^{\beta_\omega}, \quad H(\eta) = \frac{\beta_\omega}{a(\eta)\eta}, \quad \beta_\omega \equiv \frac{2}{1 + 3\omega},
\end{equation}
where $\eta_i = \beta_\omega / H_i$.

\subsection{Numerical infrastructure}

We implement the numerical scheme in the open-source numerical-relativity code \textsc{GRChombo}~\cite{Clough:2015sqa,Radia:2021smk,Andrade:2021rbd}, which employs the \textsc{Chombo} library for block-structured adaptive mesh refinement (AMR)~\cite{Adams:2015kgr}. Specifically, we incorporate the flux-conservative fluid scheme described above into the existing BSSN evolution infrastructure. We differentiate the BSSN variables using fourth-order finite-difference stencils and update the fluid conserved variables through numerical fluxes at cell faces. We integrate the coupled system in time using the method of lines and the classical fourth-order Runge--Kutta scheme. We do not apply Kreiss--Oliger dissipation in the simulations presented in this work.

Unless otherwise specified, we adopt the following fiducial initial-data parameters:
\begin{equation}
    H_i = 50, \quad k = \frac{10}{\sqrt{6}}, \quad L = 1, \quad r_W = 0.8L.
\end{equation}
Here, $H_i = 50$ can be understood as a choice of units. For these parameters and $\omega = 1 / 3$, we have $r_m = \sqrt{6} / 10$, $t_i = 0.01$, and $t_H = M_H = 1.5$.

We evolve the fiducial simulations on the octant $0 \leq x, y, z \leq L$ and impose reflection conditions at both boundaries in each Cartesian direction. For the reflection-symmetric configurations considered here, this setup is equivalent to a periodic tiling of a full box with side length $2L$, while reducing the computational cost by a factor of approximately eight. To test this equivalence, we also perform simulations on the full periodic domain using the same physical resolution and AMR criteria, as described in Sec.~\ref{sec:boundary-condition-tests}.

The base grid contains $64^3$ cells on the octant. We use five additional AMR levels with a refinement ratio of two, yielding the finest grid spacing $\Delta x_\mathrm{min} = L / 2048$. We refine the grid according to the gradients of the conformal factor $\chi$ and the Eulerian energy density $E$. The gauge conditions and corresponding time-step prescriptions are discussed in Sec.~\ref{sec:cosmological-gauges}.

For the long post-formation accretion runs in Sec.~\ref{sec:mass-growth}, we double the half-side length to $L = 2$ and retain $r_W = 0.8L$. We simultaneously double the number of base-grid cells in each direction, so that the physical resolution and AMR criteria are unchanged from the fiducial setup.

We locate marginally outer trapped surfaces using the apparent-horizon finder implemented in \textsc{GRChombo}. For the nonspinning spherical configurations considered here, we calculate the black-hole mass from the apparent-horizon area $A_\mathrm{AH}$ as
\begin{equation}
    M_\mathrm{BH} = \sqrt{\frac{A_\mathrm{AH}}{16\pi}}.
\end{equation}

\section{Cosmological gauge choices and time stepping}\label{sec:cosmological-gauges}

\subsection{Cosmic-time slicing and the cosmologically scaled Gamma-driver}

The moving-puncture gauge combines $1 + \log$ slicing with a hyperbolic Gamma-driver shift and is widely used to simulate black-hole spacetimes without excision~\cite{Alcubierre:2002kk,vanMeter:2006vi}. The collapse of the lapse prevents the time slices from reaching the physical singularity, whereas the dynamical shift alleviates the resulting coordinate stretching. However, a direct application of the standard $1 + \log$ condition to an expanding spacetime may be problematic. On an expanding FLRW background, we have $K = -3H < 0$ and the standard condition therefore drives the lapse away from unity even in the homogeneous limit (see Ref.~\cite{Aurrekoetxea:2024ypv} and references therein). Background- or mean-curvature-subtracted variants of $1+\log$ slicing have been used in cosmological numerical-relativity simulations~\cite{Zilhao:2012bb,Yoo:2013yea,Giblin:2019nuv}:
\begin{equation}
    \left(\partial_t - \beta^i \partial_i\right) \alpha = -2\alpha \left(K - \langle K\rangle\right),
    \label{eq:cosmological_lapse0}
\end{equation}
where $\langle K\rangle$ is the spatial average of $K$ over the simulation domain. This prescription admits $\alpha = 1$ in the homogeneous limit, such that the coordinate time $t$ coincides with cosmic time in the weakly perturbed region. By contrast, in a collapsing region, the local deviation $K - \langle K\rangle$ drives the collapse of the lapse and preserves the lapse-collapse mechanism of $1 + \log$ slicing. In this work, since the background is an analytical FLRW solution, we can use the mean curvature of the reference FLRW solution $K_b = -3H(t)$ instead of the spatial average $\langle K\rangle$ in Eq.~\eqref{eq:cosmological_lapse0}:
\begin{equation}
    \left(\partial_t - \beta^i \partial_i\right) \alpha = -2\alpha \left(K - K_b\right).
    \label{eq:cosmological_lapse}
\end{equation}

Although this slicing is well suited to the physical problem, long cosmological evolutions remain computationally expensive if the time step is kept proportional to the comoving grid spacing. In particular, evolving superhorizon initial data to horizon entry can involve a substantial increase in the scale factor, during which a fixed time step becomes increasingly conservative. To illustrate this issue, we consider a spatially flat FLRW region with a vanishing shift,
\begin{equation}
    \mathrm{d}s^2 = -\mathrm{d}t^2 + a^2(t) \delta_{ij} \, \mathrm{d}x^i \mathrm{d}x^j.
\end{equation}
Since the proper-time and proper-distance intervals are $\mathrm{d}\tau = \mathrm{d}t$ and $\mathrm{d}\ell = a\,\mathrm{d}x$, respectively, a signal propagating at physical speed $v \equiv \mathrm{d}\ell/\mathrm{d}\tau$ with respect to an Eulerian observer has the coordinate speed
\begin{equation}
    \frac{\mathrm{d}x}{\mathrm{d}t} = \frac{v}{a(t)}.
\end{equation}
In particular, $v = 1$ for the physical light cone, whereas $v = c_\mathrm{s} = 1 / \sqrt{3}$ for sound waves in a radiation fluid. For numerical stability, the corresponding CFL condition
\begin{equation}
    \mathcal{C}_\mathrm{phys} = \frac{v}{a(t)} \frac{\Delta t}{\Delta x} \lesssim Q,
\end{equation}
must be satisfied. Here, $Q$ is a constant of order $1$ that depends on the time integrator and spatial discretization, and $\mathcal{C}_\mathrm{phys}$ is called the CFL number. Consequently, choosing $\Delta t \propto a(t) \Delta x$ keeps the CFL number of the physical modes approximately constant and avoids an increasingly conservative time step. Since the conformal time is defined through $\mathrm{d}\eta = \mathrm{d}t / a$, this prescription yields the finite-step relation $\Delta\eta \simeq \Delta t / a = \mathrm{const.}$, although the evolution coordinate in the present construction remains cosmic time.

The lapse gauge mode obeys the same cosmological scaling. On an FLRW background with $\alpha = 1$ and $\beta^i = 0$, the local frozen-coefficient approximation to Eq.~\eqref{eq:cosmological_lapse}, together with the principal part $\partial_t K \simeq -a^{-2}\Delta \alpha$, gives $\partial_t^2\delta\alpha \simeq 2 a^{-2}\Delta \delta\alpha$ and therefore the coordinate speed $v_\alpha = \sqrt{2} / a$~\cite{Gundlach:2006tw}. The variable-step prescription thus captures the increase in the allowed time step without changing the time coordinate.

However, the standard Gamma-driver obstructs this prescription because the characteristic speeds of its shift gauge modes do not inherit the factor $a^{-1}$. To demonstrate this behavior, we write the Gamma-driver equations used in \textsc{GRChombo} as
\begin{subequations}\label{eq:gamma_driver}
    \begin{align}
        \partial_t\beta^i &= b B^i, \\
        \partial_t B^i &= \partial_t \tilde{\Gamma}^i - \eta_\mathrm{GD} B^i,
    \end{align}
\end{subequations}
where $B^i$ is an auxiliary variable, $b$ controls the response of the driver, and $\eta_\mathrm{GD} > 0$ is the damping rate. The following argument is a local frozen-coefficient analysis in the asymptotic FLRW region, rather than a complete characteristic decomposition of the nonlinear strong-field system.

On a spatially flat FLRW background, the conformal BSSN variables satisfy
\begin{equation}
    \tilde{\gamma}_{ij} = \delta_{ij}, \quad \beta^i = B^i = \tilde{\Gamma}^i = 0.
\end{equation}
The shift-dependent principal part of the BSSN equation for the conformal connection is
\begin{equation}
    \partial_t \tilde{\Gamma}^i \simeq \Delta \beta^i + \frac{1}{3}\partial^i\partial_j\beta^j, \quad \Delta \equiv \delta^{jk}\partial_j\partial_k.
    \label{eq:gamma_principal_part}
\end{equation}
The derivatives in Eq.~\eqref{eq:gamma_principal_part} are contracted with the conformal metric $\tilde{\gamma}_{ij} = \delta_{ij}$ rather than the physical metric $\gamma_{ij} = a^2 \delta_{ij}$ and therefore carry no factor of $a^{-2}$. Combining the linearized equations while retaining the lower-order damping term yields
\begin{equation}
    \partial_t^2\beta^i + \eta_\mathrm{GD}\partial_t\beta^i \simeq b \left(\Delta \beta^i + \frac{1}{3}\partial^i\partial_j\beta^j\right).
\end{equation}
For a Fourier mode proportional to $\exp[i(k_j x^j - \omega t)]$, we decompose the shift into a transverse component satisfying $k_i\beta_\mathrm{T}^i = 0$ and a longitudinal component satisfying $\beta_\mathrm{L}^i \propto k^i$. The corresponding dispersion relations are $\omega^2 + i\eta_\mathrm{GD}\omega = b k^2$ and $\omega^2 + i\eta_\mathrm{GD}\omega = 4b k^2 / 3$, respectively. The damping term affects the decay of the gauge modes but does not modify their principal characteristic speeds, which are
\begin{equation}
    v_\mathrm{T} = \sqrt{b}, \quad v_\mathrm{L} = \sqrt{\frac{4b}{3}}.
    \label{eq:shift_characteristic_speeds}
\end{equation}
Thus, the standard choice $b = 3 / 4$ gives unit coordinate speed for the fastest, longitudinal gauge mode on an asymptotically flat background. In an FLRW region, however, this speed remains unity rather than decreasing as $a^{-1}$. The associated wave CFL number is
\begin{equation}
    \mathcal{C}_\beta = \sqrt{\frac{4b}{3}} \frac{\Delta t}{\Delta x},
    \label{eq:shift_wave_cfl}
\end{equation}
which results in a global time-step restriction.

The damping term introduces a second, logically distinct restriction. For a spatially homogeneous perturbation, the spatial-derivative terms vanish, and the second of Eqs.~\eqref{eq:gamma_driver} reduces to
\begin{equation}
    \partial_t B^i = -\eta_\mathrm{GD} B^i.
\end{equation}
This equation is an ordinary differential equation, and it introduces a time step restriction~\cite{Schnetter:2010cz}:
\begin{equation}
    \mathcal{C}_\eta \equiv \eta_\mathrm{GD} \Delta t \lesssim Q',
    \label{eq:shift_damping_cfl}
\end{equation}
where $Q'$ depends on the time integrator. Unlike the wave CFL condition, this stability bound is independent of the spatial resolution. In the context of mesh refinement, Schnetter emphasized this resolution-independent time-step restriction of the BSSN Gamma-driver: the larger time steps on coarse levels can violate the damping bound even when the wave CFL condition is satisfied~\cite{Schnetter:2010cz}. He proposed a spatially varying damping coefficient $\eta_\mathrm{GD}(\boldsymbol{x})$ that decreases with distance from the black hole to avoid this violation. In the present problem, the same mechanism arises in the time domain because we want to increase $\Delta t$ as the Universe expands at fixed resolution.

Equations~\eqref{eq:shift_wave_cfl} and~\eqref{eq:shift_damping_cfl} motivate the following cosmologically scaled coefficients:
\begin{subequations}\label{eq:scaled_gamma_coefficients}
    \begin{align}
        b(t) &= b_0 \left[\frac{a_i}{a(t)}\right]^2, \\
        \eta_\mathrm{GD}(t) &= \eta_{\mathrm{GD}, 0} \frac{a_i}{a(t)}.
    \end{align}
\end{subequations}
In this work, we set $a_i = 1$, $b_0 = 3 / 4$, and $\eta_{\mathrm{GD}, 0} = 1$. With this scaling, both dimensionless stability parameters remain constant in time when $\Delta t \propto a(t) \Delta x$:
\begin{equation}
    \mathcal{C}_\beta \propto \frac{a_i}{a} \frac{\Delta t}{\Delta x}, \quad \mathcal{C}_\eta \propto \frac{a_i}{a} \Delta t.
\end{equation}
The scaling $b \propto a^{-2}$ follows directly from the wave CFL condition because the shift characteristic speed is proportional to $\sqrt{b}$. The choice $\eta_\mathrm{GD} \propto a^{-1}$ is the temporal cosmological analogue of Schnetter's damping argument and keeps $\eta_\mathrm{GD}\Delta t$ constant. This choice is a sufficient gauge prescription that preserves the stability margin of the explicit integration, rather than a unique consequence of the Einstein equations. We evaluate the coefficients in Eq.~\eqref{eq:scaled_gamma_coefficients} at each Runge--Kutta substep.

The decreasing coefficients do not imply that the driver becomes ineffective on the expanding background. Since the background proper distance is $\mathrm{d}\ell = a\,\mathrm{d}x$, the gauge-mode propagation speed measured with respect to this proper distance is $a v_\mathrm{L}$ and remains constant under the scaling in Eq.~\eqref{eq:scaled_gamma_coefficients}. For a Fourier mode with a fixed comoving wave number, our scaling choice also keeps the damping-to-frequency ratio constant because both the principal gauge frequency and $\eta_\mathrm{GD}$ scale as $a^{-1}$. This far-field argument alone does not guarantee that the scaled driver reproduces every strong-field property of the standard moving-puncture gauge near a newly formed black hole, nor does it replace direct stability tests of the coupled nonlinear system. We therefore test its strong-field behavior, constraint preservation, and convergence in Sec.~\ref{sec:results} and Appendix~\ref{app:convergence}.

For completeness, the time dependence of $b$ contributes a lower-order term when we combine the first-order driver into a second-order equation:
\begin{equation}
    \partial_t^2\beta^i + \left(\eta_\mathrm{GD} - \frac{\dot b}{b}\right)\partial_t\beta^i \simeq b \left(\Delta \beta^i + \frac{1}{3}\partial^i\partial_j\beta^j\right).
\end{equation}
Since $\dot b / b = -2H$, the effective damping in this equation is $\eta_\mathrm{GD} + 2H$. This additional expansion damping does not modify the characteristic speeds in Eq.~\eqref{eq:shift_characteristic_speeds}.

For the cosmologically scaled gauge, we choose the time step on the coarsest level according to
\begin{equation}
    \Delta t_0 = \min\left[C_\mathrm{CFL} a(t)\Delta x_0, \frac{C_H}{H(t)}\right],
\end{equation}
where $C_\mathrm{CFL} = 0.25$ and $C_H = 0.02$. The second term limits the fractional background expansion during one coarse-level step and controls the corresponding temporal truncation error in the background and source terms~\cite{Yoo:2020lmg}. It mainly reduces the time step at early times when the expansion is rapid and the typical time scale is $1/H$. With a refinement ratio of two and the Berger--Oliger time subcycling used here, the finer levels take $\Delta t_\ell = \Delta t_0 / 2^\ell$, such that $\Delta t_\ell / \Delta x_\ell$ is the same on every level.

\subsection{Conformal-time slicing}

An alternative way to remove the expansion-induced separation between the physical and coordinate propagation speeds is to use conformal time as the evolution coordinate.\footnote{A different lapse gauge condition based on the e-folding number has independently been considered for PBH formation in a matter-dominated Universe~\cite{Yoo:2026frk}.} Conformal-time evolution has previously been employed in cosmological numerical relativity. Both Daverio et al.~\cite{Daverio:2019gql} and Adshead et al.~\cite{Adshead:2023mvt} use Bona--Mass\'o lapse prescriptions of the form $(\partial_\eta - \beta^i\partial_i)\alpha = -\alpha^2 K / 3$, giving $\alpha = a$ on the FLRW background. The former adopt vanishing shift in cosmological N-body simulations and relate their gauge choice to a constant background CFL factor, whereas the latter use a standard hyperbolic Gamma-driver in fully relativistic preheating simulations. The underlying benefit of adapting the evolution time to the cosmological expansion is shared with the prescriptions considered here.

To distinguish the gauge variables associated with the two time coordinates, we temporarily write the cosmic-time lapse and shift as $\alpha_t$ and $\beta_t^i$, and their conformal-time counterparts as $\alpha_\eta$ and $\beta_\eta^i$. Under the homogeneous time reparametrization $\mathrm{d}t = a\,\mathrm{d}\eta$, the $3+1$ line element retains its form provided that
\begin{equation}
    \alpha_\eta = a \alpha_t, \quad \beta_\eta^i = a \beta_t^i.
    \label{eq:conformal-gauge-transformation}
\end{equation}
The BSSN and hydrodynamic equations retain their standard $3+1$ form when they are evolved using $\alpha_\eta$, $\beta_\eta^i$, and the coordinate time $\eta$; the change of time coordinate is therefore implemented through the gauge variables rather than by separately rescaling every evolution equation.

Applying Eq.~\eqref{eq:conformal-gauge-transformation} to the cosmic-time slicing condition in Eq.~\eqref{eq:cosmological_lapse} gives
\begin{equation}
    \left(\partial_\eta - \beta_\eta^i \partial_i\right)\alpha_\eta = \mathcal{H}\alpha_\eta - 2a \alpha_\eta\left(K - K_b\right),
    \label{eq:conformal-lapse}
\end{equation}
where a prime denotes a derivative with respect to $\eta$, $\mathcal{H} \equiv a' / a = aH$ is the conformal Hubble rate, and $K_b = -3H = -3\mathcal{H} / a$. The first term on the right-hand side accounts for the background growth of the conformal-time lapse. In the homogeneous limit, Eq.~\eqref{eq:conformal-lapse} gives $\alpha_\eta' = \mathcal{H}\alpha_\eta$ and hence $\alpha_\eta = a$ for $\alpha_{\eta,i} = a_i = 1$. Equation~\eqref{eq:conformal-lapse} drives the normalized lapse $\alpha_\eta/a$ downward where $K - K_b > 0$, thereby retaining the singularity-avoidance mechanism of the cosmic-time condition. By comparison, the lapse condition used in Refs.~\cite{Daverio:2019gql,Adshead:2023mvt} decreases $\alpha_\eta$ where $K > 0$. Our tests in Sec.~\ref{sec:results} demonstrate that our prescription supports black-hole formation and long-term post-formation evolution in the radiation-fluid configurations considered here. We leave a direct comparison of the numerical robustness of the two lapse prescriptions to future work.

We evaluate the background quantities in Eq.~\eqref{eq:conformal-lapse} analytically at every Runge--Kutta substep. With conformal time and comoving spatial coordinates, the coordinate speeds of light and sound are constant on the FLRW background. The lapse mode is likewise expansion independent: combining the principal parts of Eq.~\eqref{eq:conformal-lapse} and the $K$ equation gives $\partial_\eta^2\delta\alpha_\eta \simeq 2\Delta\delta\alpha_\eta$, up to lower-order background terms, and therefore $v_\alpha = \sqrt{2}$. We can consequently use the standard Gamma-driver with constant coefficients,
\begin{subequations}\label{eq:conformal-gamma-driver}
    \begin{align}
        \partial_\eta\beta_\eta^i &= b B_\eta^i, \\
        \partial_\eta B_\eta^i &= \partial_\eta\tilde{\Gamma}^i - \eta_\mathrm{GD} B_\eta^i.
    \end{align}
\end{subequations}
We use $b = b_0 = 3 / 4$ and $\eta_\mathrm{GD} = \eta_{\mathrm{GD},0} = 1$, so that the fastest longitudinal shift mode has unit coordinate speed and the damping timescale is constant in conformal time.

The relationship to the cosmologically scaled cosmic-time driver follows directly from Eq.~\eqref{eq:conformal-gauge-transformation}. With the spatial coordinates unchanged, $\partial_\eta = a\partial_t$, and the spatial metric and its conformal connection functions are unchanged on the same hypersurface. Defining $B^i(t,\boldsymbol{x}) \equiv B_\eta^i(\eta(t),\boldsymbol{x})$ without rescaling the auxiliary variable gives
\begin{equation}
    \partial_\eta\beta_\eta^i = a^2\left(\partial_t\beta_t^i + H\beta_t^i\right).
\end{equation}
The exact cosmic-time form of Eq.~\eqref{eq:conformal-gamma-driver} is therefore
\begin{equation}
    \begin{aligned}
        \partial_t\beta_t^i &= \frac{b_0}{a^2}B^i - H\beta_t^i, \\
        \partial_t B^i &= \partial_t\tilde{\Gamma}^i - \frac{\eta_{\mathrm{GD},0}}{a}B^i.
    \end{aligned}
    \label{eq:conformal-driver-cosmic-time}
\end{equation}
For $a_i = 1$, Eq.~\eqref{eq:conformal-driver-cosmic-time} has the same coefficient scalings as Eqs.~\eqref{eq:gamma_driver} and~\eqref{eq:scaled_gamma_coefficients}, but includes the additional term $-H\beta_t^i$ in the shift equation. Omitting this term defines our choice of spatial gauge and it is not an approximation based on assuming that the term is negligible. The additional term contains no spatial derivatives and does not change the principal shift characteristic speeds, $v_\mathrm{T} = \sqrt{b_0}/a$ and $v_\mathrm{L} = \sqrt{4b_0/3}/a$, in the asymptotic FLRW analysis. Thus, both prescriptions have the same expansion scaling of the shift wave CFL limit, although their lower-order gauge dynamics differ.

The coarsest conformal-time step is chosen as
\begin{equation}
    \Delta\eta_0 = \min\left[C_\mathrm{CFL}\Delta x_0, \frac{C_H}{\mathcal{H}(\eta)}\right],
\end{equation}
with the same $C_\mathrm{CFL} = 0.25$ and $C_H = 0.02$ as in the cosmic-time runs. The first term is the ordinary CFL limit for the constant conformal-time characteristic speeds, whereas the second limits the fractional background expansion per step at early times.

The lapse conditions are related by an exact time reparametrization, while the shift prescriptions are closely related but differ by the term identified above. Nevertheless, both constructions keep the relevant physical and gauge CFL numbers approximately constant in their respective time coordinates. We use the conformal-time evolution as a consistency and robustness check using a different evolution time coordinate and a related spatial gauge prescription. We retain cosmic time for the main threshold, critical mass scaling, and post-formation accretion results because it facilitates direct comparison with previous PBH calculations and expresses the accretion fits directly in cosmic-time quantities. These are practical conveniences rather than physical capabilities unavailable in conformal time.

\section{Numerical results}\label{sec:results}

\subsection{Gauge performance and robustness}

\begin{figure*}[htbp]
    \centering
    \includegraphics[width=0.98\textwidth]{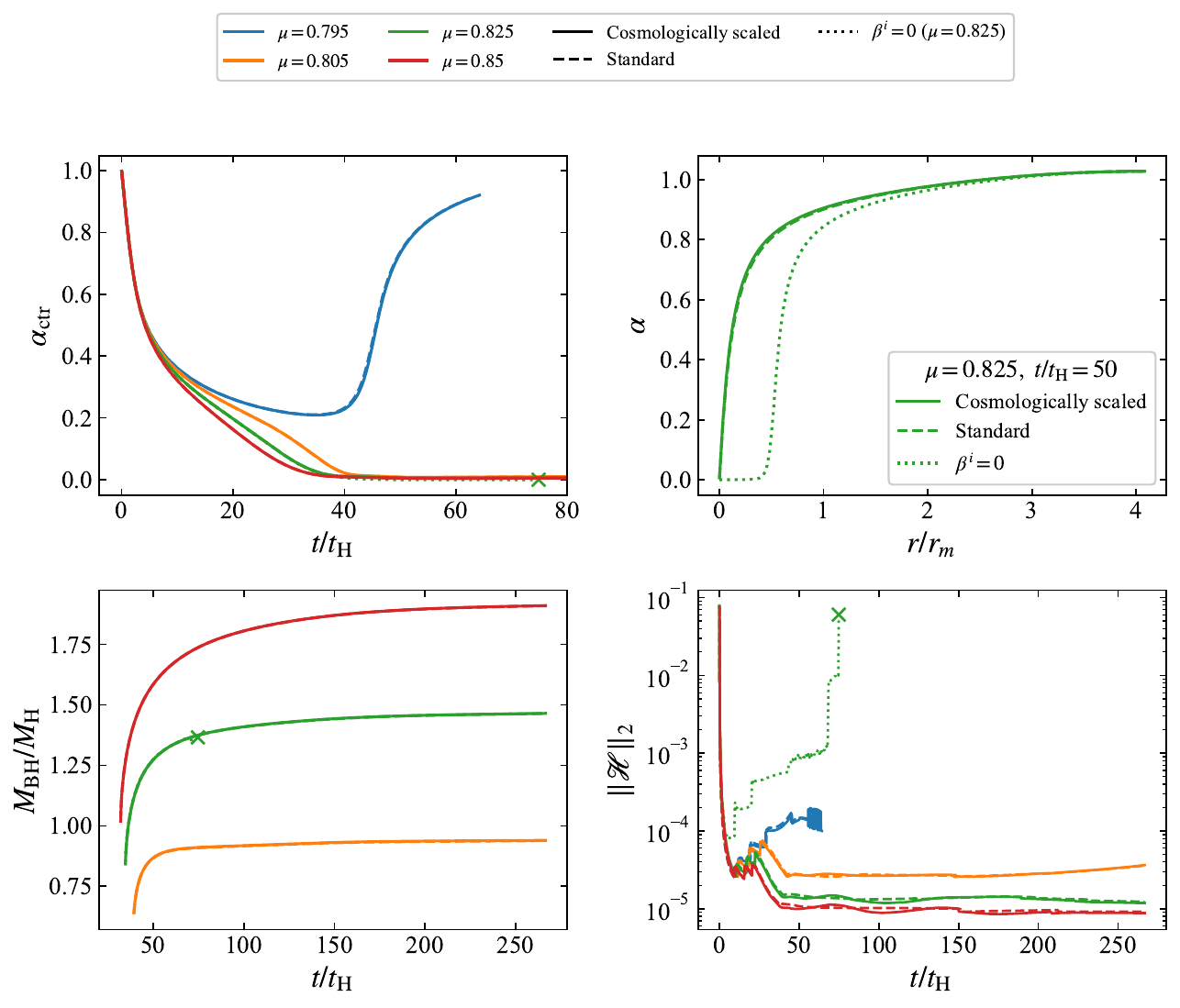}
    \caption{Gauge comparison. The upper-left panel shows the central lapse for $\mu = 0.795$, $0.805$, $0.825$, and $0.85$, and the upper-right panel compares lapse lineouts for the three $\mu = 0.825$ gauges at $t / t_\mathrm{H} = 50$. The lower-left panel compares apparent-horizon masses for the three supercritical amplitudes, while the lower-right panel shows the $L^2$ norm of the Hamiltonian constraint for all four amplitudes. Colors identify $\mu$ as indicated in the legend. Solid, dashed, and dotted curves denote the scaled driver, standard driver, and vanishing shift, respectively; the vanishing-shift comparison is performed only for $\mu = 0.825$. Crosses mark its last valid samples.}
    \label{fig:gauge-comparison}
\end{figure*}

We first test whether the cosmological scaling of the Gamma-driver changes the collapse dynamics or the resulting black-hole properties, and how the increased time step affects the numerical stability and computational efficiency. Fig.~\ref{fig:gauge-comparison} compares the scaled and standard drivers for a subcritical configuration with $\mu = 0.795$ and supercritical configurations with $\mu = 0.805$, $0.825$, and $0.85$. Runs with the standard Gamma-driver use $\Delta t_0 = \min(C_\mathrm{CFL} \Delta x_0, C_H/H(t))$, because its gauge mode does not redshift with the expansion. For $\mu = 0.825$, we additionally perform a comparison run in which we evolve the lapse with the same cosmological $1 + \log$ condition while freezing the shift to $\beta^i = 0$. This vanishing-shift run uses the same time-step prescription as the scaled driver, since there is no shift CFL condition to consider. The central lapse in the upper-left panel decreases and subsequently bounces in the subcritical case, whereas it collapses and remains small after black-hole formation in the supercritical cases. The scaled and standard drivers give nearly identical histories for all four amplitudes. The vanishing-shift run also agrees with the dynamical-shift results until $t / t_\mathrm{H} = 74.87$, after which the simulation breaks down. Crosses mark the last valid vanishing-shift samples. The lapse lineouts at $t / t_\mathrm{H} = 50$ in the upper-right panel illustrate the developing coordinate problem: the scaled and standard Gamma-drivers give nearly overlapping profiles, whereas freezing the shift produces a much broader region of collapsed lapse and stronger coordinate stretching.

The coordinate evolution is gauge dependent, as expected, but the apparent-horizon masses in the lower-left panel are robust. All mass comparisons, including the vanishing-shift run before its breakdown, agree well over their common intervals. The agreement of the horizon masses before the crash supports interpreting the differences between the corresponding field profiles as gauge rather than physical, while the subsequent breakdown shows that lapse collapse alone is insufficient for a robust long post-formation evolution. These results also demonstrate that the decreasing Gamma-driver coefficients do not become ineffective in the strong-field region, and that the scaled driver retains the singularity-avoidance and coordinate-stretching mitigation properties of the standard moving-puncture gauge.

The lower-right panel of Fig.~\ref{fig:gauge-comparison} shows the $L^2$ norm of the Hamiltonian constraint. The relatively large values of $\Vert\mathscr{H}\Vert_2$ at early times are AMR-induced numerical transients, arising primarily from non-constraint-preserving coarse-to-fine interpolation and mismatched finite-difference truncation errors across refinement levels. These transients rapidly decay as the evolution relaxes. For all four amplitudes, the scaled and standard drivers produce constraint histories of the same magnitude, including through lapse collapse and apparent-horizon formation in the supercritical cases. By contrast, the norm in the vanishing-shift run begins a sustained growth after $t / t_\mathrm{H} \simeq 60$ and reaches $6.04 \times 10^{-2}$ at $t / t_\mathrm{H} = 74.87$, after which the simulation breaks down.

The computational advantage of the scaled driver follows from the time-step histories. With the standard driver, the nonredshifting longitudinal gauge mode enforces the constant cap $\Delta t / t_\mathrm{H} \simeq 2.60 \times 10^{-3}$. By contrast, the time step of the scaled driver grows with the expansion and reaches $\Delta t / t_\mathrm{H} \simeq 0.520$ at the end of the fiducial run, a factor of approximately $200$ larger. We use the $\mu = 0.825$ evolutions as a direct step-count comparison: reaching $t / t_\mathrm{H} \simeq 266.8$ requires $1092$ coarse-level advances with the scaled driver and $102436$ with the standard driver, a reduction by a factor of approximately $94$. This comparison isolates the algorithmic gain and does not rely on machine-dependent wall-clock timings. In our runs, the wall-clock time is also reduced by a factor of approximately $92$, which is consistent with the step-count comparison.

\begin{figure*}[htbp]
    \centering
    \includegraphics[width=0.98\textwidth]{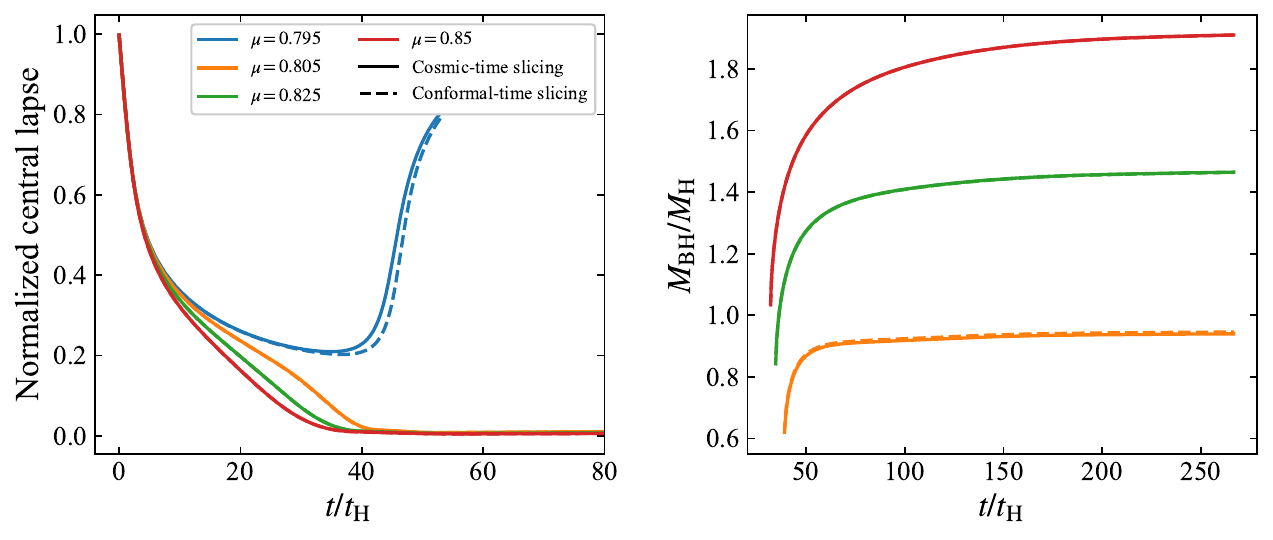}
    \caption{Comparison of the cosmologically scaled cosmic-time gauge and the conformal-time gauge with the standard Gamma-driver for $\mu = 0.795$, $0.805$, $0.825$, and $0.85$. Conformal-time diagnostics are plotted against the background-mapped cosmic time. The left panel compares the cosmic-time central lapse $\alpha_t$ with the normalized conformal-time lapse $\alpha_\eta / a$ through $t / t_\mathrm{H} = 80$. The right panel compares the apparent-horizon masses for the three supercritical amplitudes through the long post-formation evolution. Colors identify $\mu$, while solid and dashed curves denote cosmic- and conformal-time slicing, respectively.}
    \label{fig:conformal-time-comparison}
\end{figure*}

We next compare the cosmologically scaled cosmic-time gauge with the conformal-time gauge. For the radiation background, the corresponding elapsed cosmic time is $t - t_i = \int_{\eta_i}^\eta a(\bar\eta)\,\mathrm{d}\bar\eta = (\eta - \eta_i) + H_i(\eta - \eta_i)^2 / 2$. We use this background relation to express diagnostics from the conformal-time runs in the same cosmic-time normalization as the fiducial results; it does not imply that the strong-field slices of the two gauges coincide pointwise. Fig.~\ref{fig:conformal-time-comparison} compares the two constructions for $\mu = 0.795$, $0.805$, $0.825$, and $0.85$. In the three supercritical cases, the cosmic-time lapse $\alpha_t$ and the normalized conformal-time lapse $\alpha_\eta / a$ track one another closely through collapse. For the subcritical case, both gauges describe contraction followed by dispersal, although a phase difference during the lapse bounce produces a larger discrepancy. These coordinate-field differences are expected because the full gauge systems are not related by a pure time reparametrization. More importantly, the apparent-horizon mass histories agree well through the evolution.

For the representative $\mu = 0.825$ pair, the two prescriptions also have comparable costs. The cosmic-time and conformal-time runs require $1092$ and $1090$ coarse-level advances, respectively, and their recorded wall-clock times are nearly equal. The latter comparison is machine dependent, but together with the nearly identical advance counts it shows that changing the evolution coordinate does not introduce a substantial computational penalty. The agreement of the horizon masses provides a consistency and robustness check of the implementation and results, while the remainder of this section uses the cosmic-time construction for the reasons given above.

\subsection{Periodic domain and black-hole interior regularization}\label{sec:boundary-condition-tests}

\begin{figure*}[htbp]
    \centering
    \includegraphics[width=0.98\textwidth]{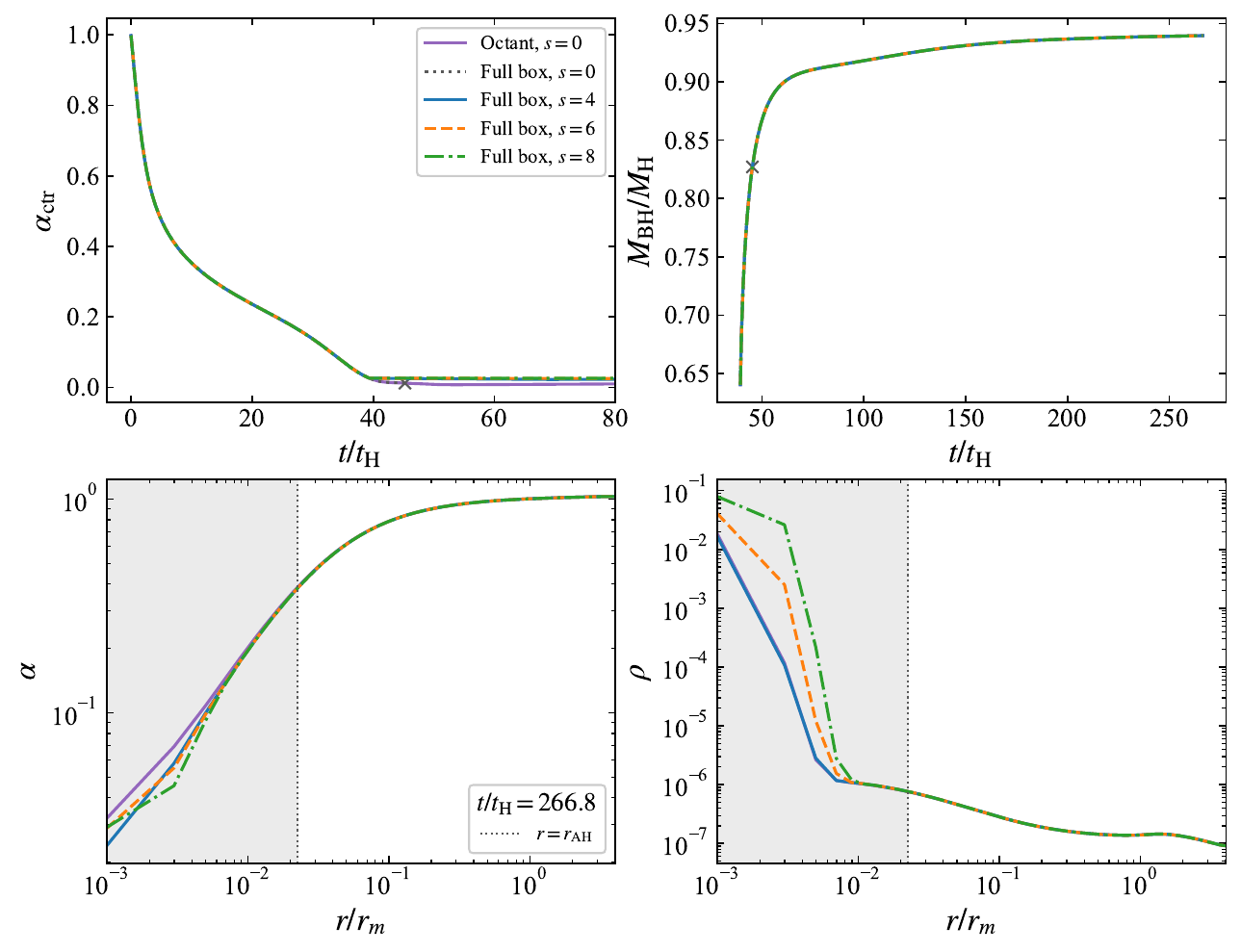}
    \caption{Boundary-condition and black-hole-interior regularization tests for $\mu = 0.805$. The top row compares the central lapse and apparent-horizon mass in the reflective octant without regularization with those in the full periodic box for $s = 0$, $4$, $6$, and $8$. Crosses mark the last valid point of the unregularized full-box run with $s = 0$. The bottom row compares terminal lineouts of the lapse and proper energy density for the octant and the three regularized full-box runs at $t / t_\mathrm{H} = 266.8$. The shaded region is inside the apparent horizon, and the vertical dotted line marks its coordinate radius.}
    \label{fig:boundary-comparison}
\end{figure*}

The reflection conditions used in our fiducial octant simulations enforce the discrete symmetries of the spherical initial data and reduce the computational cost. To relax the enforced symmetries, we repeat the $\mu = 0.805$ simulation on the full periodic box with the same resolution. As shown in the top row of Fig.~\ref{fig:boundary-comparison}, the central lapse and apparent-horizon mass agree with the octant result until the full-box evolution fails at $t / t_\mathrm{H} \simeq 45.3$. This agreement demonstrates that the reflection boundaries reproduce the full periodic solution for the spherical configuration over their common valid interval.

However, the failure of the full-box run needs to be understood. In our tests, we find that the breakdown of the full-box simulation occurs mainly after an apparent horizon has formed. We attribute it to the increasingly steep and underresolved fields in the black-hole interior. The octant symmetry suppresses nonspherical truncation-error modes across the coordinate planes, whereas the full box permits such modes to develop inside the horizon. We therefore regard the different termination times as a numerical interior effect rather than a physical dependence on the boundary conditions.

Inspired by the black-hole excision procedure in the \textsc{COSMOS} code~\cite{Yoo:2021fxs,Yoo:2024lhp}, in which the evolution of a central region inside the apparent horizon is frozen, we introduce an excision-like interior regularization after the first apparent horizon is found to extend the full periodic-box evolution. This procedure does not remove grid cells or impose an inner boundary. Instead, at every subsequent right-hand-side evaluation, we multiply the right-hand side of every evolved variable by
\begin{equation}
    w_\mathrm{reg}(r) = \begin{cases} \exp\left[-s\left(1 - r^2 / r_\mathrm{reg}^2\right)^p\right], & r < r_\mathrm{reg}, \\ 1, & r \geq r_\mathrm{reg}, \end{cases}
\end{equation}
where $r$ is the coordinate distance from the apparent-horizon center, $s$ is the regularization strength, $p$ is the power, $r_\mathrm{reg} = f_\mathrm{reg} r_\mathrm{AH,min}$ is the regularization radius, and $r_\mathrm{AH,min}$ is the current minimum coordinate radius of the apparent horizon. Setting $s = 0$ gives $w_\mathrm{reg} = 1$ everywhere and therefore disables the interior regularization. We use $f_\mathrm{reg} = 0.5$ and $p = 4$, with $s = 4$, $6$, and $8$ in the regularized full-box comparison runs. The procedure therefore progressively slows the coordinate evolution only well inside the apparent horizon. Outside that region, the equations and numerical update are unaltered. Although physical signals cannot propagate from this region to the exterior, gauge and constraint modes need not obey the physical light cone. We therefore validate the procedure empirically rather than relying only on causal separation.

As shown in Fig.~\ref{fig:boundary-comparison}, all three regularized full-box runs reach $t / t_\mathrm{H} = 266.8$, and their apparent-horizon masses all agree with the octant result. The bottom row of Fig.~\ref{fig:boundary-comparison} compares terminal lineouts of the lapse and proper energy density. Varying $s$ from $4$ to $6$ or $8$ visibly changes the fields deep inside the apparent horizon, as intended, but has a negligible effect on the exterior fields. The simultaneous agreement of the horizon mass and exterior fields provides the practical validation required for the long post-formation evolutions using the full periodic box with interior regularization. We note that the interior regularization is used only in the full-box robustness tests of this subsection and is not employed in the fiducial octant simulations used for the threshold, critical-scaling, and accretion results below.

\subsection{Collapse threshold and critical mass scaling}

We next identify the collapse threshold between dispersal and black-hole formation by varying the curvature amplitude $\mu$. The central lapse in the upper-left panel of Fig.~\ref{fig:gauge-comparison} provides a convenient dynamical indicator. For $\mu = 0.795$, the lapse initially decreases as the overdensity contracts, reaches a nonzero minimum, and then returns toward unity as pressure gradients disperse the perturbation. For $\mu = 0.805$, $0.825$, and $0.85$, it instead continues to collapse, and an apparent horizon is subsequently found. The lapse behavior is gauge dependent and is therefore not used by itself to define black-hole formation; the decisive diagnostic is the appearance of a marginally outer trapped surface.

To identify the threshold and fit the critical mass scaling, we use an additional refinement level and halve the AMR refinement thresholds relative to the fiducial setup, giving $\Delta x_\mathrm{min} = L / 4096$. At this resolution, an amplitude scan finds that the threshold lies between $\mu = 0.79578$ and $0.79580$. A scan with twice the base-grid resolution and half the AMR tagging thresholds shifts the interval to approximately $0.79583 < \mu_c < 0.79585$, as discussed in Appendix~\ref{app:critical-exponent-tests}. These results refine the earlier three-dimensional result $0.795 < \mu_c < 0.805$ obtained with the \textsc{COSMOS} code~\cite{Yoo:2020lmg} and the more recent three-dimensional result $0.7 < \mu_c < 0.8$~\cite{Baumgarte:2026igz} for the same spherical Gaussian curvature profile, although the latter study initialized the perturbation at horizon entry using a different constraint-satisfying prescription. The baseline threshold interval is also close to the spherically symmetric Misner--Sharp result $\mu_c \simeq 0.79579 \pm 1 \times 10^{-5}$ for the same curvature profile~\cite{Ning:2025yvj}. The agreement between the three-dimensional BSSN and spherically symmetric Misner--Sharp results provides a validation of the matter evolution in our three-dimensional code.

\begin{figure}[htbp]
    \centering
    \includegraphics[width=0.95\linewidth]{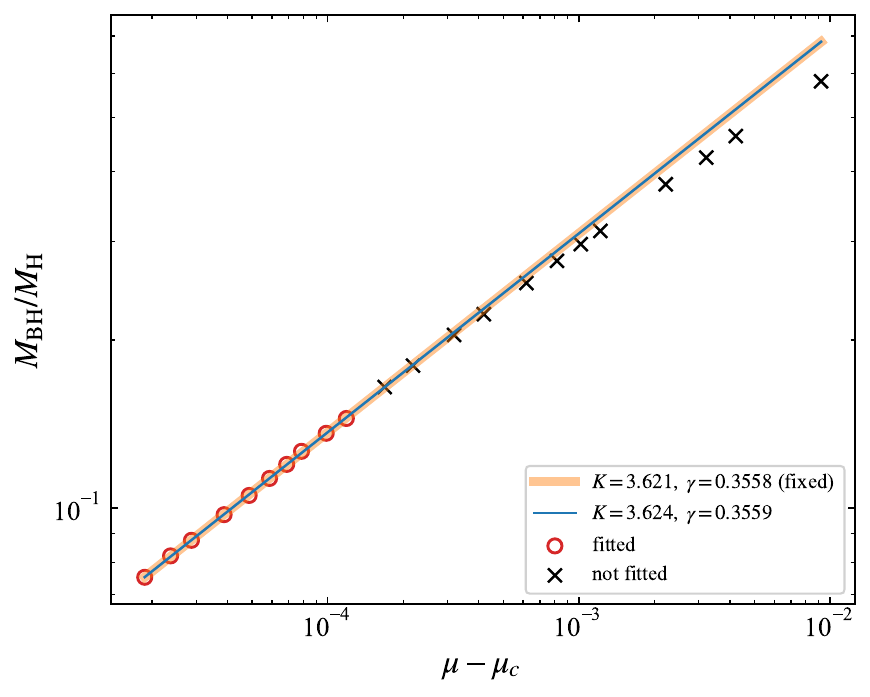}
    \caption{Apparent-horizon mass at first detection as a function of the distance from the jointly fitted collapse threshold. Open circles denote the ten nearest available supercritical points used in the fit, while crosses show farther supercritical points excluded from the fit. The thin blue line is the joint fit of $(\mathcal{K},\mu_c,\gamma)$, while the thick orange line fixes $\gamma = 0.3558$ and fits $(\mathcal{K},\mu_c)$.}
    \label{fig:critical-scaling}
\end{figure}

Near the collapse threshold, the PBH mass is expected to obey the scaling relation
\begin{equation}
    M_\mathrm{BH} = M_\mathrm{H}\mathcal{K}(\mu - \mu_c)^\gamma,
\end{equation}
where $\mathcal{K}$ depends on the initial-data family and mass normalization, whereas the critical exponent $\gamma$ depends only on the fluid equation of state within the critical-collapse regime. For a radiation fluid, previous analytical and numerical studies have found $\gamma \simeq 0.3558$~\cite{Evans:1994pj,Koike:1995jm,Neilsen:1998qc,Niemeyer:1999ak,Hawke:2002rf,Musco:2004ak,Musco:2008hv,Musco:2012au}. Fig.~\ref{fig:critical-scaling} shows the apparent-horizon mass at first detection as a function of the perturbation amplitude. Following Ref.~\cite{Yoo:2021fxs}, we perform the fit using the ten nearest available supercritical points. A joint fit of $(\mathcal{K},\mu_c,\gamma)$ gives $\mathcal{K} \simeq 3.624$, $\mu_c \simeq 0.7957813$, and $\gamma \simeq 0.3559$ (the thin blue line). Fixing the exponent to the radiation-fluid value $\gamma = 0.3558$ instead gives $\mathcal{K} \simeq 3.621$ and essentially the same value of $\mu_c$ (the thick orange line). The free- and fixed-exponent fits are nearly indistinguishable over the fitted range. By contrast, the unfitted points farther above threshold bend away from the leading power law. However, varying the fitting interval gives $0.3498 \lesssim \gamma \lesssim 0.3787$ at this resolution, while the corresponding higher-resolution fits give $0.3262 \lesssim \gamma \lesssim 0.3410$. These tests, presented in Appendix~\ref{app:critical-exponent-tests}, support approximate critical mass scaling but not a precision continuum measurement of $\gamma$.

\subsection{Post-formation mass growth}\label{sec:mass-growth}

\begin{figure}[htbp]
    \centering
    \includegraphics[width=0.98\linewidth]{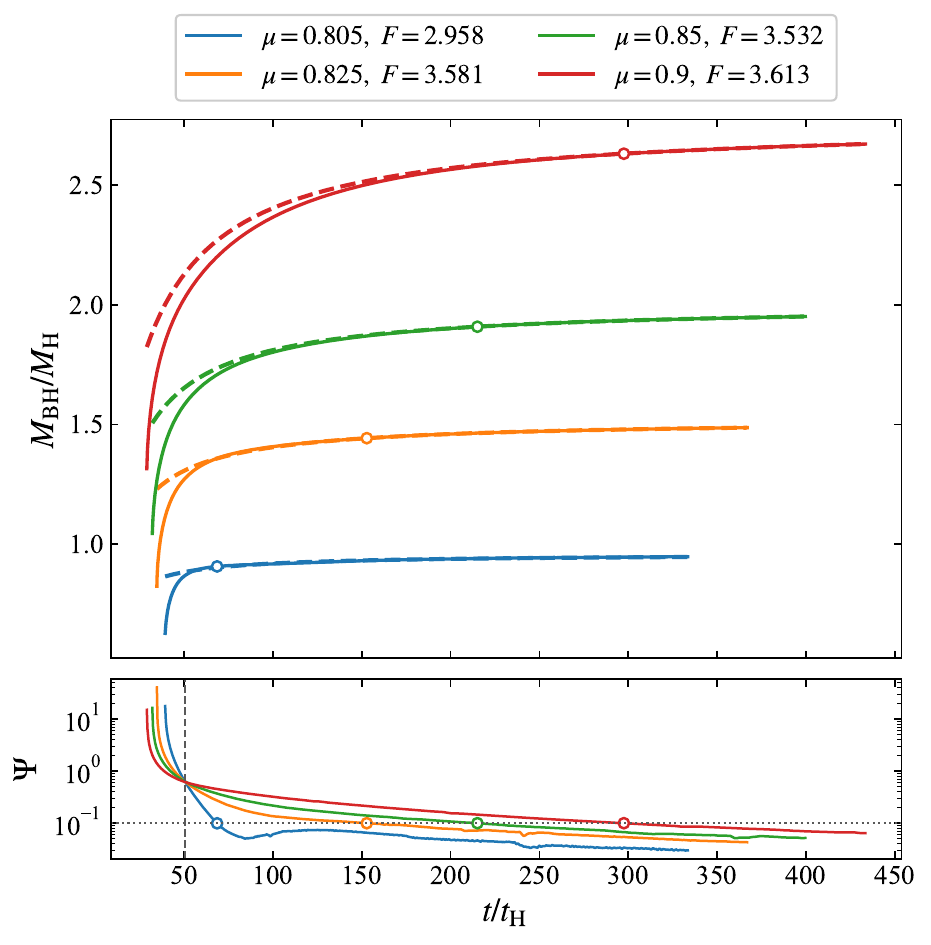}
    \caption{Post-formation black-hole mass growth and late-time accretion fits. The upper panel shows the apparent-horizon mass for four amplitudes; dashed curves are fits to Eq.~\eqref{eq:accretion-solution}, extended before the fitting interval, and open circles mark the first fitted points. The lower panel shows $\Psi$ from Eq.~\eqref{eq:psi-accretion}. The horizontal dotted line marks $\Psi = 0.1$, and the vertical dashed line marks the approximate common crossing of the four $\Psi$ curves.}
    \label{fig:mass-growth}
\end{figure}

The cosmologically scaled gauge allows us to perform long-term evolutions of post-formation black holes efficiently. We follow the apparent horizon long after formation and measure the declining accretion rate. We compare the numerical mass evolution with the late-time Zel'dovich--Novikov prescription~\cite{Zeldovich:1967lct} used in Refs.~\cite{Deng:2016vzb,Escriva:2019nsa,Yoo:2021fxs}. Approximating the exterior by a radiation-dominated FLRW background and writing the effective accretion rate as
\begin{equation}
    \dot M_\mathrm{BH} = 4\pi F R_\mathrm{BH}^2 \rho_b(t) = \frac{3F}{2}\frac{M_\mathrm{BH}^2}{t^2},
\end{equation}
where $R_\mathrm{BH} = 2M_\mathrm{BH}$, $\rho_b$ is the background energy density, and $F$ is a dimensionless effective efficiency, we obtain
\begin{equation}
    M_\mathrm{BH}(t) = \left[\frac{1}{M_\infty} + \frac{3F}{2t}\right]^{-1}.
    \label{eq:accretion-solution}
\end{equation}
Here, $M_\infty$ is the extrapolated asymptotic mass. This formula is expected to be valid only at sufficiently late times after black-hole formation. Following Ref.~\cite{Escriva:2019nsa}, we monitor
\begin{equation}
    \Psi \equiv \frac{\dot M_\mathrm{BH}}{H M_\mathrm{BH}},
    \label{eq:psi-accretion}
\end{equation}
and fit the contiguous interval from the first point satisfying $\Psi \leq 0.1$ to the end of each run. Fig.~\ref{fig:mass-growth} shows the mass histories for $\mu = 0.805$, $0.825$, $0.85$, and $0.9$ in the enlarged $L = 2$ domain. The dashed curves are the fits to Eq.~\eqref{eq:accretion-solution}, extrapolated to earlier times to display where the approximation ceases to follow the rapid post-formation growth. Open circles mark the first fitted point.

\begin{table}[htbp]
    \caption{Late-time accretion fits using the contiguous data from the first point with $\Psi \leq 0.1$ to the end of each run.}
    \label{tab:accretion-fits}
    \begin{ruledtabular}
        \begin{tabular}{ccc}
            $\mu$ & $F$ & $M_\infty / M_\mathrm{H}$ \\
            \hline
            $0.805$ & $2.958$ & $0.958$ \\
            $0.825$ & $3.581$ & $1.520$ \\
            $0.850$ & $3.532$ & $2.004$ \\
            $0.900$ & $3.613$ & $2.765$ \\
        \end{tabular}
    \end{ruledtabular}
\end{table}

The fitted efficiencies in Table~\ref{tab:accretion-fits} span $2.958 \lesssim F \lesssim 3.613$. For the three larger amplitudes, they cluster in the narrower range $3.532 \lesssim F \lesssim 3.613$, consistent with the range $3.5 \lesssim F \lesssim 3.75$ found by Escriv\`a in the Misner--Sharp simulations of Ref.~\cite{Escriva:2019nsa}; the near-threshold $\mu = 0.805$ run instead gives the smaller value $F = 2.958$. We note that the value of $F$ depends on the foliation of the time slices~\cite{Yoo:2021fxs}. Interestingly, we also observe a common crossing of the $\Psi$ curves, as reported in Ref.~\cite{Escriva:2019nsa}. The approximate common crossing occurs at $t / t_\mathrm{H} = 50.6$, where $\Psi \simeq 0.62$.

We test the influence of domain size in Appendix~\ref{app:domain-size-tests}. For $\mu = 0.805$ and $0.9$, doubling the domain size while keeping the resolution and initial curvature perturbation fixed changes the fitted $F$ by at most approximately $0.114\%$ and $M_\infty$ by at most approximately $0.0223\%$.

\section{Conclusions and discussion}\label{sec:conclusions}

In this work, we have developed a three-dimensional numerical-relativity framework for following PBH formation from an initially superhorizon curvature perturbation through long-term post-formation evolution in a radiation-dominated Universe. Specifically, we implemented a flux-conservative relativistic-fluid module in the AMR code \textsc{GRChombo}, constructed growing-mode initial data using the long-wavelength expansion, and coupled the hydrodynamic evolution to the BSSN formulation. To reduce the temporal cost of cosmological evolution, we introduced a cosmologically scaled Gamma-driver with $b \propto a^{-2}$ and $\eta_\mathrm{GD} \propto a^{-1}$, which permits $\Delta t \propto a \Delta x$ in cosmic-time slicing while preserving the moving-puncture behavior required near the black hole. We also constructed a conformal-time version of the moving-puncture gauge as a consistency check.

The scaled driver reproduces the apparent-horizon masses and constraint behavior obtained with the standard driver while substantially reducing the computational cost of long evolutions. For the representative $\mu = 0.825$ configuration evolved to $t / t_\mathrm{H} \simeq 266.8$, the number of coarse-level advances is reduced by a factor of approximately $94$. The conformal-time calculation yields consistent horizon masses at a comparable cost. Boundary-condition, black-hole-interior regularization, resolution, and domain-size tests further support the numerical robustness of the exterior observables used in this work.

For a spherical Gaussian curvature perturbation, our three-dimensional simulations find the collapse threshold $\mu_c \simeq 0.7958$, close to the spherically symmetric Misner--Sharp result for the same profile. The near-critical fits give $0.3262 \lesssim \gamma \lesssim 0.3787$, depending on the resolution and fitting interval. Thus, the simulations exhibit approximate critical mass scaling. These results validate the three-dimensional fluid evolution and show that the framework resolves the near-critical regime. The accelerated time evolution also allows us to follow the post-formation evolution over many Hubble times and fit the late-time PBH mass growth to the Zel'dovich--Novikov accretion prescription. The fitted effective efficiencies are broadly consistent with previous spherically symmetric results, although they are foliation-dependent. To our knowledge, this work presents the first determination of both the critical mass scaling relation and the post-formation accretion law from three-dimensional PBH formation simulations.

The present application is restricted to a spherical Gaussian perturbation in a radiation fluid, but the numerical framework can be extended to other initial-data families and barotropic perfect fluids with other constant values of $\omega$. Natural extensions include nonspherical profiles characterized by ellipticity and prolateness, for which we plan to investigate their effects on the PBH formation threshold, critical mass scaling, and spin. The AMR infrastructure also makes the code well suited to multi-peak curvature perturbations, which may produce multiple PBHs or lead to cooperative collapse. More broadly, the cosmologically scaled Gamma-driver and the conformal-time gauge may be useful for other cosmological simulations, including massless scalar-field collapse in an expanding Universe~\cite{Yoo:2018pda,Yoo:2021fxs,Padilla:2026ieh}, kination-dominated cosmologies~\cite{Cheng:2025eas,Joana:2026myf}, and quartic scalar-field collapse~\cite{Milligan:2025zbu,Padilla:2025bkv}. However, for an oscillating massive scalar field, the intrinsic oscillation frequency introduces an expansion-independent timescale that would prevent the time step from continuing to grow proportionally to the scale factor.

\begin{acknowledgments}
    This work is supported by the National Key Research and Development Program of China Grant No. 2021YFC2203004, No. 2021YFA0718304, and No. 2020YFC2201501, the National Natural Science Foundation of China Grants No. 12422502, No. 12547110, No. 12588101, No. 12235019, and No. 12447101, and the Science Research Grants from the China Manned Space Project with No. CMS-CSST-2025-A01. C.Y. is supported by JSPS KAKENHI Grant Numbers JP25K07281 and JP24K07027.
\end{acknowledgments}

\appendix

\section{Fluid evolution scheme}\label{app:fluid}

In this Appendix, we summarize the hydrodynamic evolution scheme adapted from the formulation used in the \textsc{COSMOS} code~\cite{Escriva:2024lmm}. Since we adopt a barotropic equation of state without separating the rest-mass and internal-energy contributions, we evolve only the densitized energy and momentum variables $(\mathcal{E}, \mathcal{S}_i)$ rather than an additional baryon rest-mass density.

\subsection{Conserved-to-primitive recovery}

After each time-integration substep, we first recover the Eulerian variables from their densitized counterparts as
\begin{equation}
    E = \frac{\mathcal{E}}{\sqrt{\gamma}}, \quad S_i = \frac{\mathcal{S}_i}{\sqrt{\gamma}}.
\end{equation}
Defining $S^2 \equiv \gamma^{ij} S_i S_j$, Eq.~\eqref{eq:fluid_projections} implies $S^2 = (E + p)^2 v^2$. For the linear barotropic equation of state $p = \omega \rho$, eliminating the Lorentz factor gives
\begin{equation}
    \omega \rho^2 + (1 - \omega)E\rho - (E^2 - S^2) = 0.
\end{equation}
For $\omega > 0$, the positive-energy solution is
\begin{equation}
    \rho = \frac{-(1 - \omega)E + \sqrt{(1 - \omega)^2 E^2 + 4\omega(E^2 - S^2)}}{2\omega}.
\end{equation}
For $\omega = 0$, we instead have $\rho = (E^2 - S^2) / E$. The remaining primitive variables then follow from
\begin{equation}
    p = \omega \rho, \quad v^i = \frac{S^i}{E + p}, \quad W = (1 - v^2)^{-1/2}.
\end{equation}
Therefore, the conserved-to-primitive conversion does not require an iterative root finder.

\subsection{MUSCL reconstruction and numerical flux}

We reconstruct the primitive variables component by component using the monotonic upstream-centered scheme for conservation laws (MUSCL) with a generalized minmod limiter~\cite{Kurganov:2000ovy,Shibata:2005jv}. Consider one coordinate direction and suppress the transverse grid indices. For any primitive variable $q \in \{\rho, v^1, v^2, v^3\}$, we define the forward and backward differences at grid point $I$ by
\begin{equation}
    (\Delta_+ q)_I = q_{I + 1} - q_I, \quad (\Delta_- q)_I = q_I - q_{I - 1},
\end{equation}
and the limited differences by
\begin{equation}
    \begin{aligned}
        (\overline{\Delta}_+ q)_I &= \operatorname{minmod}\left[(\Delta_+ q)_I, b_\mathrm{lim}(\Delta_- q)_I\right], \\
        (\overline{\Delta}_- q)_I &= \operatorname{minmod}\left[(\Delta_- q)_I, b_\mathrm{lim}(\Delta_+ q)_I\right].
    \end{aligned}
\end{equation}
Here, the minmod function is defined as
\begin{equation}
    \operatorname{minmod}(x,y) =
    \begin{cases}
        \operatorname{sgn}(x)\min(|x|,|y|), & xy > 0, \\
        0, & xy \leq 0.
    \end{cases}
\end{equation}
The left and right states at the face $I + 1/2$ are reconstructed according to
\begin{equation}
    \begin{aligned}
        q_{I + 1/2}^{L} &= q_I + \frac{1 - \kappa}{4}(\overline{\Delta}_- q)_I + \frac{1 + \kappa}{4}(\overline{\Delta}_+ q)_I, \\
        q_{I + 1/2}^{R} &= q_{I + 1} - \frac{1 - \kappa}{4}(\overline{\Delta}_+ q)_{I + 1} - \frac{1 + \kappa}{4}(\overline{\Delta}_- q)_{I + 1}.
    \end{aligned}
\end{equation}
We set $\kappa = 1 / 3$ and choose the limiter parameter $b_\mathrm{lim} = 2$. The geometric and gauge variables are interpolated to the same face using fourth-order midpoint interpolation. We then convert the reconstructed primitive states to the densitized conserved states $\boldsymbol{U}_{L,R} = (\mathcal{E}, \mathcal{S}_1, \mathcal{S}_2, \mathcal{S}_3)_{L,R}^{\mathsf{T}}$ using the common face-centered geometry.

For a face normal to the $x^k$ direction, the components of the physical flux $\boldsymbol{F}^{(k)} = (F_{\mathcal{E}}^{(k)}, F_{\mathcal{S}_1}^{(k)}, F_{\mathcal{S}_2}^{(k)}, F_{\mathcal{S}_3}^{(k)})^{\mathsf{T}}$ are
\begin{equation}
    \begin{aligned}
        F_{\mathcal{E}}^{(k)} &= \mathcal{E}(\alpha v^k - \beta^k) + \alpha\sqrt{\gamma}\,p v^k, \\
        F_{\mathcal{S}_m}^{(k)} &= \mathcal{S}_m(\alpha v^k - \beta^k) + \alpha\sqrt{\gamma}\,p\delta_m{}^k, \quad m = 1, 2, 3.
    \end{aligned}
\end{equation}
We combine the left and right states using the local Lax--Friedrichs flux,
\begin{equation}
    \widehat{\boldsymbol{F}}^{(k)} = \frac{1}{2} \left[\boldsymbol{F}^{(k)}(\boldsymbol{U}_L) + \boldsymbol{F}^{(k)}(\boldsymbol{U}_R) - a_\star^{(k)}(\boldsymbol{U}_R - \boldsymbol{U}_L)\right].
\end{equation}
The coordinate characteristic speeds are
\begin{equation}
    \begin{aligned}
        \lambda_0^{(k)} &= \alpha v^k - \beta^k, \\
        \lambda_\pm^{(k)} &= -\beta^k + \frac{\alpha}{1 - \omega v^2} \left[(1 - \omega)v^k \pm \sqrt{\mathcal{D}_k}\right],
    \end{aligned}
\end{equation}
where $\mathcal{D}_k \equiv \omega(1 - v^2) \left[\gamma^{kk}(1 - \omega v^2) - (1 - \omega)(v^k)^2\right]$. We evaluate the local propagation speed from both reconstructed states as
\begin{equation}
    a_\star^{(k)} = \max_{A \in \{L, R\}} \left\{\left|\lambda_{0,A}^{(k)}\right|, \left|\lambda_{+,A}^{(k)}\right|, \left|\lambda_{-,A}^{(k)}\right|\right\}.
\end{equation}
Only the extremal eigenvalues are required; the method does not rely on a complete characteristic decomposition or an exact Riemann solver.

The semi-discrete flux-conservative evolution equation for a cell-centered conserved state is
\begin{equation}
    \frac{\mathrm{d}\boldsymbol{U}_{\boldsymbol{I}}}{\mathrm{d}t} = -\sum_{k = 1}^{3}\frac{\widehat{\boldsymbol{F}}^{(k)}_{\boldsymbol{I} + \boldsymbol{e}_k / 2} - \widehat{\boldsymbol{F}}^{(k)}_{\boldsymbol{I} - \boldsymbol{e}_k / 2}}{\Delta x_k} + \boldsymbol{\mathcal{Q}}_{\boldsymbol{I}},
\end{equation}
where $\boldsymbol{e}_k$ is the unit grid vector in the $x^k$ direction and $\boldsymbol{\mathcal{Q}} = (\mathcal{Q}_E, \mathcal{Q}_{S_i})$ collects the cell-centered geometric source terms. We recompute the primitive variables, reconstructed states, fluxes, and source terms at every Runge--Kutta substep. At coarse--fine AMR interfaces, each refinement level evaluates its fluxes independently.

\section{Numerical tests}\label{app:convergence}

In this Appendix, we examine the convergence behavior, domain-size dependence of the accretion results, and sensitivity of the near-critical results to resolution and fitting interval.

\subsection{Convergence tests}

We perform a resolution study for the representative supercritical configuration with $\mu = 0.805$ in the fiducial $L = 1$ domain. Denoting the number of base-grid cells along each octant direction by $N$, the low-, medium-, and high-resolution runs use $N = 32$, $64$, and $128$, respectively, with the same five additional AMR levels. The corresponding finest grid spacings are $\Delta x_\mathrm{min} = L / 1024$, $L / 2048$, and $L / 4096$. To maintain comparable refinement coverage as the grid is successively doubled, we simultaneously halve the tagging thresholds for the gradients of $\chi$ and $E$.

\begin{figure*}[htbp]
    \centering
    \includegraphics[width=0.98\textwidth]{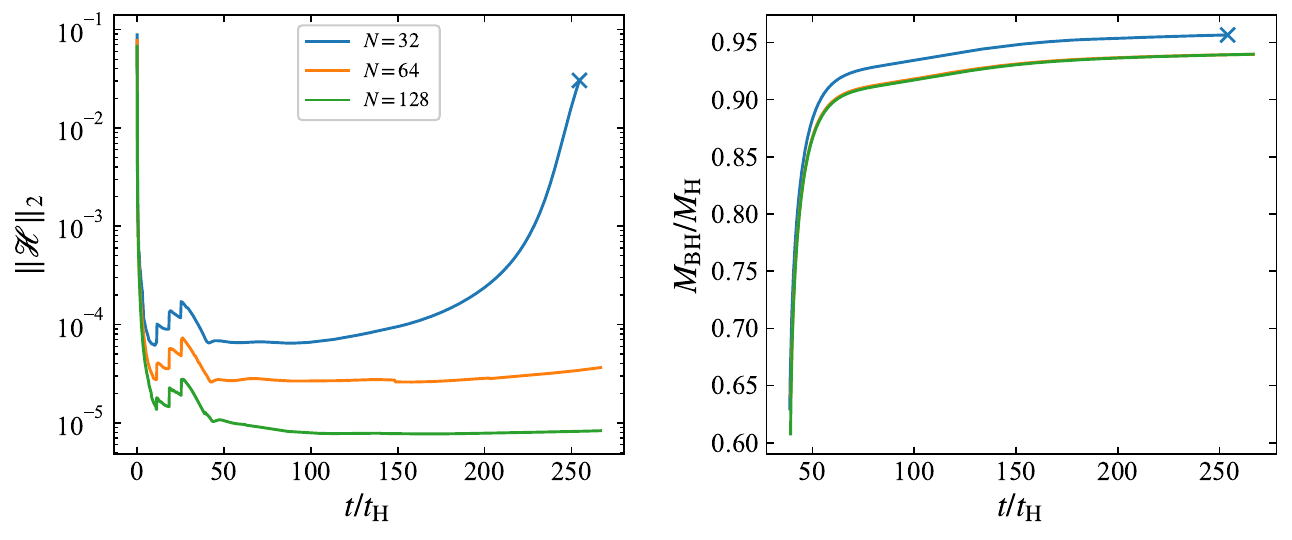}
    \caption{Resolution comparison for the $\mu = 0.805$ fiducial simulation. The left panel shows the $L^2$ norm of the Hamiltonian constraint, and the right panel shows the apparent-horizon mass. The three curves correspond to octant base-grid sizes $N = 32$, $64$, and $128$; the AMR tagging thresholds are reduced by a factor of two between successive resolutions. Crosses mark the last valid low-resolution samples before numerical breakdown.}
    \label{fig:convergence-comparison}
\end{figure*}

Fig.~\ref{fig:convergence-comparison} compares the $L^2$ norm of the Hamiltonian constraint and the apparent-horizon mass. The constraint norm decreases systematically with increasing resolution over the evolution; in particular, the late growth visible at low resolution is strongly suppressed at medium resolution and absent at high resolution over the simulated interval. The low-resolution evolution eventually breaks down, and crosses mark its last valid samples. The black-hole mass histories agree closely for the medium and high resolutions, while the low-resolution result shows visible deviations.

\begin{figure*}[htbp]
    \centering
    \includegraphics[width=0.98\textwidth]{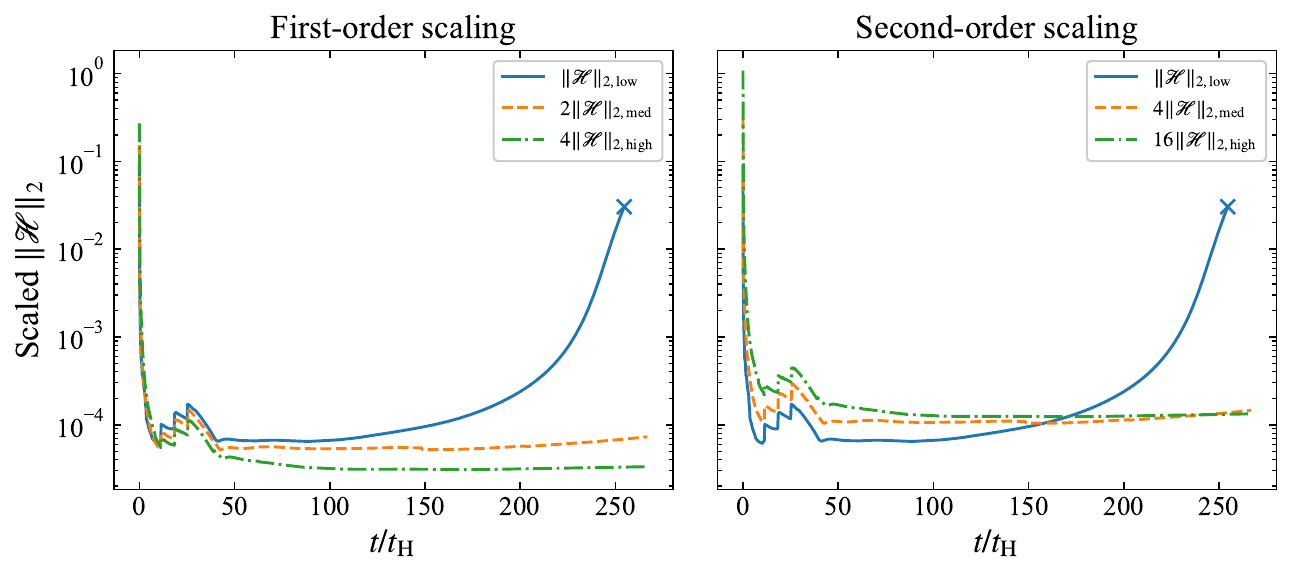}
    \caption{Scaled Hamiltonian-constraint norms for the $\mu = 0.805$, $L = 1$ resolution study. The low-, medium-, and high-resolution runs have octant base grids $N = 32$, $64$, and $128$, respectively. The left panel applies the first-order scaling factors $(1,2,4)$, while the right panel applies the second-order factors $(1,4,16)$. Crosses mark the last finite low-resolution diagnostic sample before numerical breakdown.}
    \label{fig:hamiltonian-scaled-convergence}
\end{figure*}

Since the continuum Hamiltonian constraint vanishes, a residual norm dominated by an error proportional to the $p$th power of the grid spacing would satisfy
\begin{equation}
    \Vert\mathscr{H}\Vert_{2,\mathrm{low}} \simeq 2^p\Vert\mathscr{H}\Vert_{2,\mathrm{med}} \simeq 4^p\Vert\mathscr{H}\Vert_{2,\mathrm{high}}.
\end{equation}
We therefore compare the norms scaled by $(1,2,4)$ for first-order convergence and $(1,4,16)$ for second-order convergence in Fig.~\ref{fig:hamiltonian-scaled-convergence}.

After the initial adjustment and before the low-resolution constraint violation grows substantially, approximately $20 \lesssim t/t_\mathrm{H} \lesssim 100$, the norm decreases by factors between roughly $2$ and $4$ upon each doubling of resolution. This behavior is consistent with an empirical convergence order between one and two for the Hamiltonian-constraint norm over this interval. At later times, the medium- and high-resolution norms approach second-order scaling, whereas the low-resolution evolution deteriorates and eventually develops non-finite diagnostics. The initial adjustment and the late low-resolution deterioration are not described by a single convergence order.

\subsection{Domain-size tests}\label{app:domain-size-tests}

Periodic images and disturbances propagating through the finite domain can affect the surrounding fluid at late times, so the domain-size dependence of the inferred accretion parameters should be tested directly. We repeat the $\mu = 0.805$ and $0.9$ accretion simulations in a larger domain. These configurations span the amplitudes used in our accretion analysis, with $\mu = 0.9$ also giving the largest black-hole mass and the longest evolution in that sample.

For each configuration, we increase the half-side length from $L = 2$ to $L = 4$, so that the corresponding full periodic box has side length $4$ or $8$, respectively. We increase the octant base grid from $128^3$ to $256^3$ cells to retain the same resolution, while keeping other settings unchanged. We also keep the initial curvature perturbation and its truncation window fixed: the taper starts at $r = 1.6$, and the curvature perturbation vanishes for $r \geq 2$ in both domains. Thus, the window function in the $L = 4$ runs retains the cutoff radius of the $L = 2$ runs rather than scaling with the larger domain. The larger calculation adds an initially FLRW exterior without changing the central perturbation. Each new run is evolved from the initial slice to the same final time as its $L = 2$ counterpart.

\begin{figure}[htbp]
    \centering
    \includegraphics[width=0.98\linewidth]{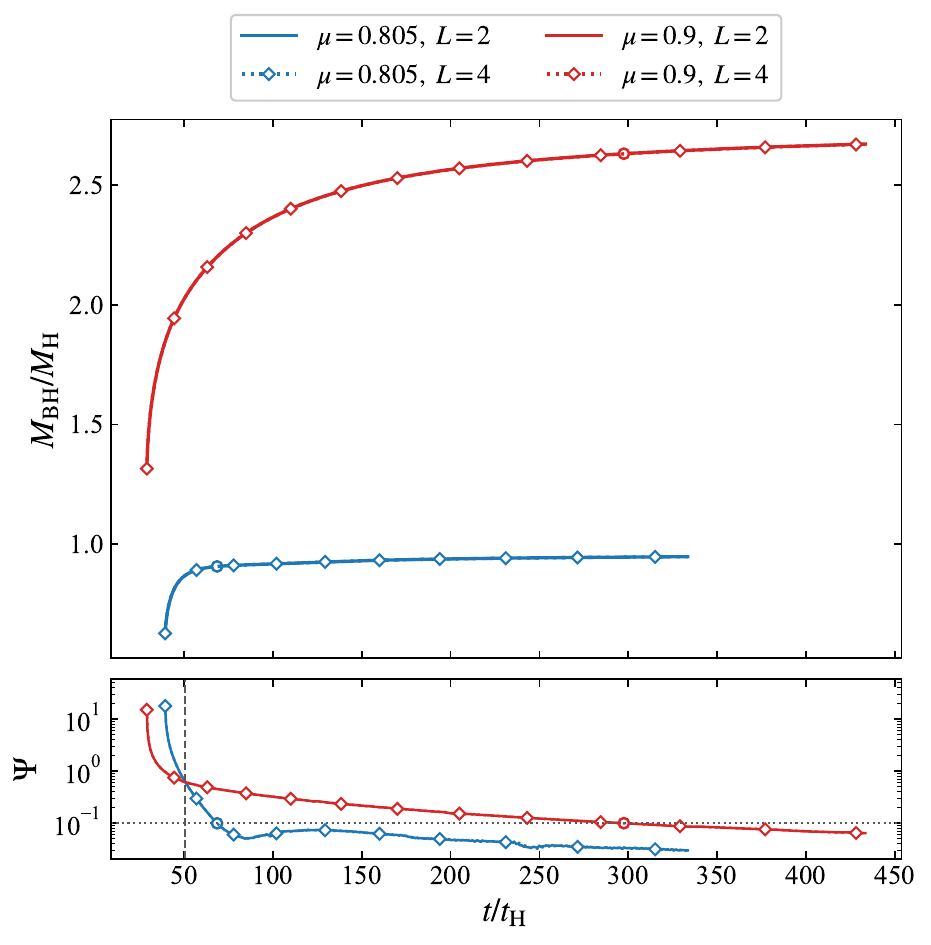}
    \caption{Domain-size comparison for $\mu = 0.805$ (blue) and $0.9$ (red). The upper panel shows the apparent-horizon mass, and the lower panel shows $\Psi$. Solid curves denote $L = 2$; dotted curves with sparsely placed open diamonds denote $L = 4$. Open circles mark the first fitted points. The horizontal dotted line marks $\Psi = 0.1$, and the vertical dashed line reproduces the reference crossing time in Fig.~\ref{fig:mass-growth}.}
    \label{fig:domain-size-comparison}
\end{figure}

Figure~\ref{fig:domain-size-comparison} compares the apparent-horizon mass and $\Psi$ for the two domain sizes. The mass histories and the evolution of $\Psi$ closely agree for both amplitudes. Defining the symmetric relative mass difference as $\delta_M(t) = 2|M_2(t) - M_4(t)|/[M_2(t) + M_4(t)]$, where the subscripts denote $L$, we find maximum differences of approximately $0.00464\%$ for $\mu = 0.805$ and $0.0221\%$ for $\mu = 0.9$ over their respective common mass histories.

We fit both mass histories with the accretion model in Eq.~\eqref{eq:accretion-solution}. For each amplitude, the two mass histories are fitted over the same cosmic-time interval selected using $\Psi \leq 0.1$. Table~\ref{tab:domain-size-fits} gives the results. Relative to the $L = 2$ values, increasing the domain changes $F$ by approximately $0.114\%$ for $\mu = 0.805$ and $0.0103\%$ for $\mu = 0.9$. The corresponding changes in $M_\infty$ are approximately $0.00538\%$ and $0.0223\%$. Thus, the agreement extends to the inferred accretion parameters, not only to the accumulated black-hole mass. These tests support retaining the $L = 2$ accretion results in Sec.~\ref{sec:mass-growth}.

\begin{table}[htbp]
    \caption{Accretion fits over identical time intervals for each pair of domain sizes.}
    \label{tab:domain-size-fits}
    \begin{ruledtabular}
        \begin{tabular}{cccc}
            $\mu$ & $L$ & $F$ & $M_\infty/M_\mathrm{H}$ \\
            \hline
            $0.805$ & $2$ & $2.95801$ & $0.958359$ \\
            $0.805$ & $4$ & $2.95463$ & $0.958308$ \\
            $0.900$ & $2$ & $3.61348$ & $2.765087$ \\
            $0.900$ & $4$ & $3.61310$ & $2.764469$ \\
        \end{tabular}
    \end{ruledtabular}
\end{table}

\subsection{Critical-exponent tests}\label{app:critical-exponent-tests}

We first examine the resolution dependence of the collapse threshold. We perform an additional near-threshold scan with twice the base-grid resolution and half the AMR tagging thresholds. The baseline and higher-resolution scans use octant base grids with $N = 64$ and $128$, respectively, in the same $L = 1$ domain, with six additional AMR levels in both cases. The finest grid spacings are therefore $\Delta x_\mathrm{min} = 1/4096$ and $1/8192$.

The baseline working interval, $0.79578 < \mu_c < 0.79580$, shifts to approximately $0.79583 < \mu_c < 0.79585$ at the higher resolution. Specifically, the higher-resolution $\mu = 0.79583$ run has no detected horizon and its central lapse recovers after maximum compression, whereas $\mu = 0.79585$ produces a horizon that can subsequently be tracked. The midpoints of the two working intervals differ by about $5\times10^{-5}$, which exceeds the width of the baseline interval. This demonstrates that the baseline bracket is not an estimate of the continuum accuracy.

The leading critical mass-scaling law is an asymptotic relation as $\mu \to \mu_c^+$. At fixed numerical resolution, however, moving closer to threshold produces smaller black holes and makes the collapse scales increasingly difficult to resolve accurately. Farther from threshold, corrections to the leading power law can become important. The fitting interval must therefore balance these two limitations.

To isolate fitting-interval dependence from resolution effects, we first use the baseline simulation data and repeat the fit
\begin{equation}
    \ln\left(\frac{M_\mathrm{BH}}{M_\mathrm{H}}\right) = \ln\mathcal{K} + \gamma\ln(\mu - \mu_c)
\end{equation}
with $(\mathcal{K},\mu_c,\gamma)$ all free. The baseline ten-point interval is $0.795800 \leq \mu \leq 0.795900$, which gives $\gamma = 0.3559$. We vary the fitting interval by either removing up to two points in total from its endpoints or adding the next one or two available points at its upper end. No interior points are removed. Table~\ref{tab:scaling-window-tests} summarizes the results, including the baseline fit.

\begin{table}[htbp]
    \caption{Dependence of the freely fitted critical exponent on the fitting interval at the baseline resolution. We remove up to two points from the endpoints of the ten-point sample or add the next one or two larger-amplitude points. The parameter $n$ is the number of fitted points.}
    \label{tab:scaling-window-tests}
    \begin{ruledtabular}
        \begin{tabular}{cccc}
            $\mu_\mathrm{min}$ & $\mu_\mathrm{max}$ & $n$ & $\gamma$ \\
            \hline
            $0.795800$ & $0.795900$ & $10$ & $0.3559$ \\
            $0.795805$ & $0.795900$ & $9$ & $0.3587$ \\
            $0.795810$ & $0.795900$ & $8$ & $0.3498$ \\
            $0.795800$ & $0.795880$ & $9$ & $0.3668$ \\
            $0.795800$ & $0.795860$ & $8$ & $0.3787$ \\
            $0.795805$ & $0.795880$ & $8$ & $0.3775$ \\
            $0.795800$ & $0.795950$ & $11$ & $0.3564$ \\
            $0.795800$ & $0.796000$ & $12$ & $0.3536$ \\
        \end{tabular}
    \end{ruledtabular}
\end{table}

These eight fitting intervals give $0.3498 \lesssim \gamma \lesssim 0.3787$. Thus, even at fixed resolution, modest endpoint changes shift the fitted exponent at the level of a few times $10^{-2}$. This range describes sensitivity to the specified fitting intervals. The four decimal places in the baseline best-fit value $0.3559$ should therefore not be interpreted as the accuracy of a continuum determination of the critical exponent.

Since the numerical threshold shifts at higher resolution, we translate the baseline ten amplitudes by $5\times10^{-5}$, preserving their spacings and the interval width. The higher-resolution fitted amplitudes are $\mu = 0.795850$, $0.795855$, $0.795860$, $0.795870$, $0.795880$, $0.795890$, $0.795900$, $0.795910$, $0.795930$, and $0.795950$.

\begin{figure}[htbp]
    \centering
    \includegraphics[width=0.95\linewidth]{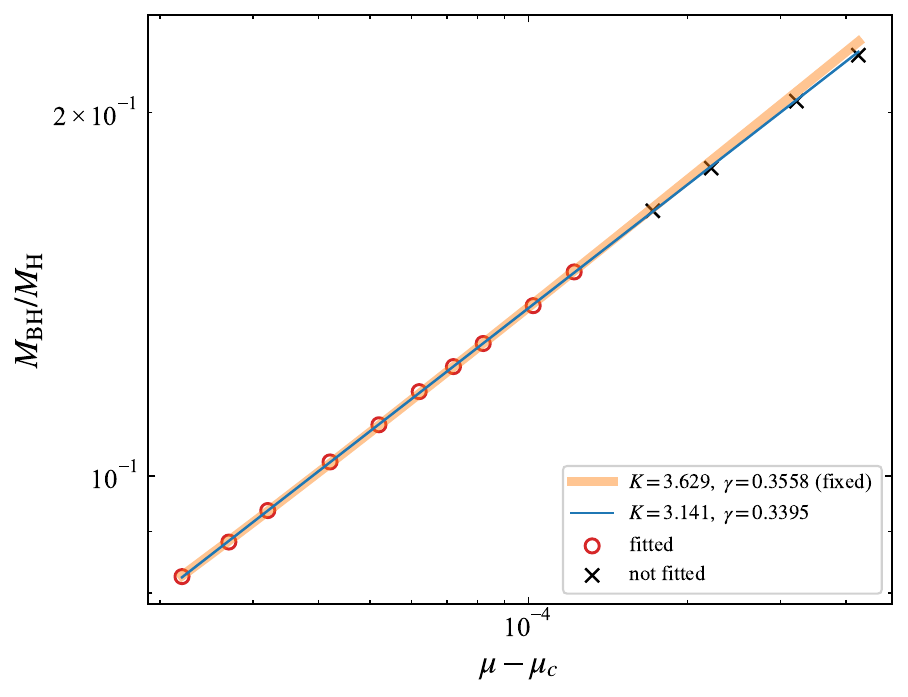}
    \caption{Critical mass-scaling fits for the higher-resolution scan with octant base grid $N = 128$ and six additional AMR levels. Open circles denote the ten fitted amplitudes. Black crosses show $\mu = 0.79600$, $0.79605$, $0.79615$, and $0.79625$, which are excluded from the displayed fits. These amplitudes are the corresponding larger points in the baseline scan translated by $5\times10^{-5}$. The thin blue curve has a free exponent, while the broad orange curve fixes $\gamma = 0.3558$.}
    \label{fig:high-resolution-scaling}
\end{figure}

Figure~\ref{fig:high-resolution-scaling} shows the higher-resolution sample and the two fits. With all three parameters free, we obtain $\mathcal{K} = 3.1411$, $\mu_c = 0.79582799$, and $\gamma = 0.33945$. Fixing $\gamma = 0.3558$ gives $\mathcal{K} = 3.6293$ and $\mu_c = 0.79582581$. The corresponding curve also closely describes the fitted data.

The full-window exponent changes from $0.3559$ in the baseline sample to $0.3395$ in this higher-resolution sample, a shift of approximately $0.0164$. This difference quantifies the change between these two finite-resolution fits. To assess the fitting-interval dependence, we use the same prescription as at the baseline resolution: we either remove up to two points in total from the endpoints or add the next one or two amplitudes, $\mu = 0.79600$ and $0.79605$, at the upper end. No interior points are removed. Table~\ref{tab:high-resolution-window-tests} lists the results.

\begin{table}[htbp]
    \caption{Fitting-interval dependence at higher resolution. We remove up to two points from the endpoints of the ten-point sample or add the next one or two larger amplitudes. All three scaling parameters are free.}
    \label{tab:high-resolution-window-tests}
    \begin{ruledtabular}
        \begin{tabular}{cccc}
            $\mu_\mathrm{min}$ & $\mu_\mathrm{max}$ & $n$ & $\gamma$ \\
            \hline
            $0.795850$ & $0.795950$ & $10$ & $0.3395$ \\
            $0.795855$ & $0.795950$ & $9$ & $0.3348$ \\
            $0.795860$ & $0.795950$ & $8$ & $0.3385$ \\
            $0.795850$ & $0.795930$ & $9$ & $0.3352$ \\
            $0.795850$ & $0.795910$ & $8$ & $0.3407$ \\
            $0.795855$ & $0.795930$ & $8$ & $0.3262$ \\
            $0.795850$ & $0.796000$ & $11$ & $0.3410$ \\
            $0.795850$ & $0.796050$ & $12$ & $0.3371$ \\
        \end{tabular}
    \end{ruledtabular}
\end{table}

These eight fitting intervals give $0.3262 \lesssim \gamma \lesssim 0.3410$. These tests exhibit a smaller endpoint sensitivity than for the baseline sample, but the fitted exponents remain systematically below the radiation-fluid value. The closer agreement with the theoretical exponent at the baseline resolution does not imply higher numerical accuracy. The exponent is fitted jointly with the threshold and normalization over a finite interval, so changes in the mass data can be partly absorbed by correlated changes in these parameters. Discretization effects and finite-window corrections can bias the effective exponent in different directions and may partially cancel at one resolution.

Taken together, these tests quantify a resolution-dependent shift of the collapse transition and changes in the effective critical exponent at the level of a few times $10^{-2}$ under variations of resolution and fitting interval. The simulations exhibit approximate critical mass scaling, but the present scans do not provide a precision continuum measurement of $\gamma$.

\bibliography{ref.bib}

\end{document}